\documentclass[journal]{IEEEtran}
\usepackage{multirow}
\usepackage[hidelinks]{hyperref}
\usepackage{amssymb,amsmath}
\usepackage{rotating}
\usepackage[ruled,vlined]{algorithm2e}
\usepackage{algorithmic}
\usepackage{bm}
\usepackage{graphicx}
\usepackage{subfigure}
\usepackage{color}
\usepackage{orcidlink}
\usepackage{multirow}
\usepackage{amsthm}
\usepackage{amsbsy}
\usepackage{booktabs}
\usepackage{multirow}
\usepackage{graphicx}
\usepackage{balance}
\usepackage{multicol} 
\usepackage{multirow}
\usepackage{booktabs}
\usepackage{enumitem} 
\usepackage{graphicx}
\usepackage{hyperref}
\usepackage[table]{xcolor} 
\usepackage{tabularx}
\usepackage{multirow}

\definecolor{burntorange}{RGB}{181,72,18}
\definecolor{custompurple}{RGB}{106, 45, 145}
\ifCLASSINFOpdf
\else
\fi

\begin{document}

\title{Beyond Single View: Cross‑Modal Graph Transformation for LLM-based Recommendation}

\author{Yibiao~Wei~\orcidlink{0009-0001-7241-9826}, Jie Zou$^{*}$~\orcidlink{0000-0002-0181-778X}, Weikang Guo~\orcidlink{0000-0003-3532-6943}, Xiao Ao~\orcidlink{0009-0009-2912-3868}, Chaoning Zhang~\orcidlink{0000-0001-6007-6099}, Yang Yang~\orcidlink{0000-0002-5070-4511},~\IEEEmembership{Senior Member,~IEEE},\\ Heng Tao Shen~\orcidlink{0000-0002-2999-2088},~\IEEEmembership{Fellow,~IEEE}.

\thanks{$^{*}$ Corresponding author.}%
\thanks{Yibiao Wei, Jie Zou, Xiao Ao, Chaoning Zhang, and Yang Yang are with the School of Computer Science and Engineering, University of Electronic Science and Technology of China, Chengdu, China (e-mail: weiyibiao12138@gmail.com; jie.zou@uestc.edu.cn; 202422080439@sta.uestc.edu.cn; chaoningzhang1990@gmail.com; yang.yang@uestc.edu.cn).}
\thanks{Weikang Guo is with the School of Management Science and Engineering, Southwestern University of Finance and Economics, Chengdu, China (e-mail: guowk@swufe.edu.cn).}
\thanks{Heng~Tao~Shen is with the School of Computer Science and Technology, Tongji University, Shanghai, China (e-mail: shenhengtao@tongji.edu.cn).}
}

\markboth{IEEE TRANSACTIONS ON MULTIMEDIA}%
{Wei \MakeLowercase{\textit{et al.}}: Cross‑Modal Graph Transformation for LLM-based Recommendation}

\maketitle

\begin{abstract}
Large Language Models (LLMs) have recently been introduced into graph-based recommendation to enhance the semantic understanding of traditional graph-based recommendation methods, typically by converting graphs into textual sequences. However, such methods still face two major challenges: (1) long input sequences exacerbate the Lost-in-the-Middle'' problem, causing the model to underutilize information in the middle of long contexts, (2) the multi-hop and multi-relational topological dependencies in graphs are linearized into sequential text, leading to semantic flattening and weakening structural modeling capacity. 
To address these limitations, we propose GTRec, a framework for multimodal LLM (MLLM)-based recommendation. 
To alleviate semantic flattening, we transform the MITG into a visual modality, which preserves global topology and complex graph dependencies beyond linearized textual sequences.
To mitigate the Lost-in-the-Middle'' problem caused by long textual inputs, we introduce Cluster-based Graph Compression (CGC), which clusters semantically similar nodes into compact descriptions while preserving the core semantics of user interests. 
Finally, we design a multimodal prompt template that integrates visualized MITG and compressed text into MLLMs, enabling joint reasoning across modalities. Experiments on three datasets show that our method achieves state-of-the-art performance, validating its effectiveness in recommendation accuracy.
\end{abstract}


\IEEEpeerreviewmaketitle

\section{Introduction}
Recommender systems improve the efficiency of information access by delivering personalized content, thereby alleviating the growing challenge of information overload. The diverse relations among users, items, and auxiliary entities in recommender systems naturally lend themselves to graph-based modeling, such as user--item interaction graphs \cite{lightgcn, CORONA,tmm2,tmm1}, knowledge graphs \cite{kgllmrec, kgrec1}, and social graphs \cite{social1,social2}. By leveraging these graph structures, graph-based recommendation methods \cite{lightgcn, NGCF, graphrec3, graphrec4} can effectively capture high-order associations and mitigate data sparsity.

Despite their effectiveness in modeling structural dependencies, traditional graph-based recommendation methods often struggle to capture complex semantics, contextual nuances, and reasoning paths that go beyond explicitly defined graph topology \cite{kgllmrec,GLRec}. Fortunately, recent advances in Large Language Models (LLMs) have demonstrated strong capabilities in semantic understanding and reasoning, inspiring a growing body of work that integrates graphs with LLMs for recommendation \cite{G-Refer,CORONA,GLRec,LLMRG,LLMREC,GaCLLM}. Representative approaches organize graph evidence from different perspectives, verbalize heterogeneous relational patterns such as meta-paths, or adopt stepwise descriptions to make graph structures more accessible to LLMs \cite{G-Refer,GLRec,GaCLLM}. Despite their differences, these methods largely rely on converting graph structures into textual forms, such as serialized triples, path descriptions, or hierarchical summaries, so that LLMs can exploit graph information through language. However, such graph-to-text paradigms still suffer from two major challenges.

\begin{figure}[tbp]
  \centering
  \begin{minipage}{.49\linewidth}
    \centering
    \includegraphics[width=\linewidth]{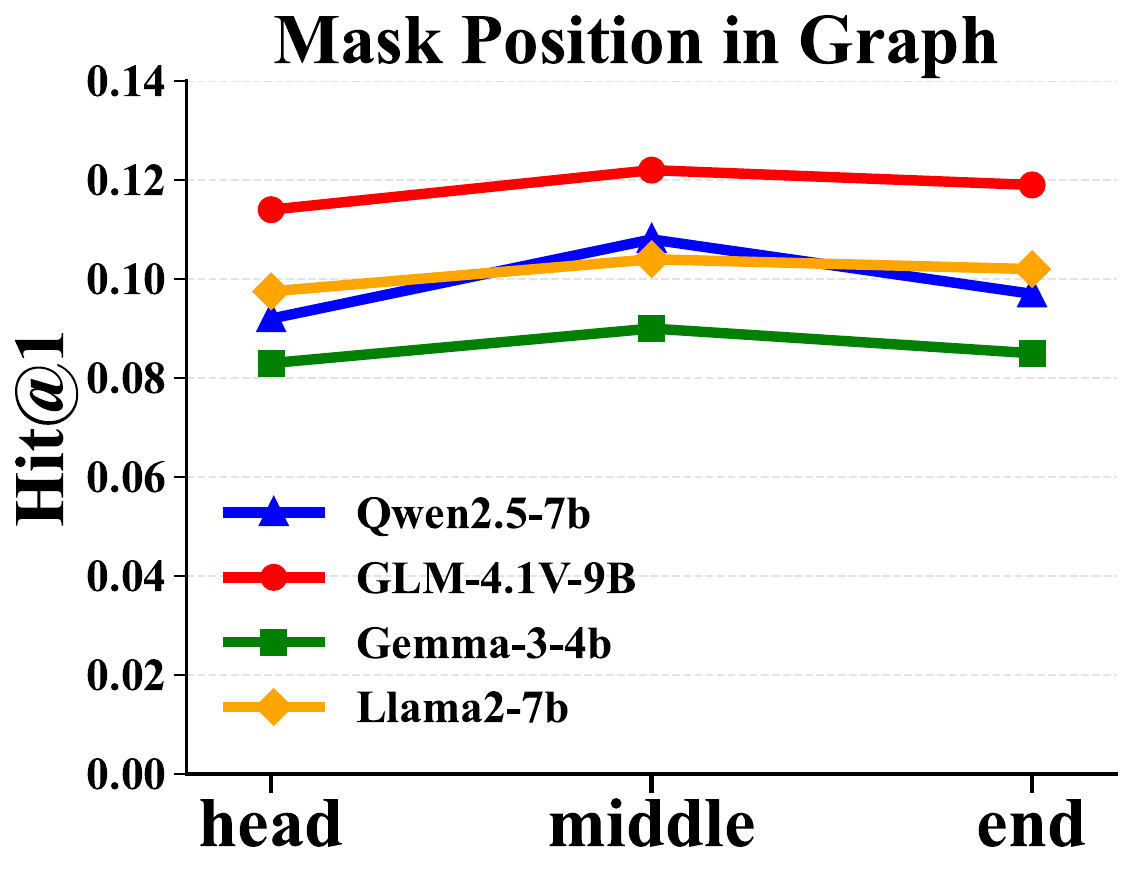}
  \end{minipage}\hfill
  \begin{minipage}{.49\linewidth}
    \centering
    \includegraphics[width=\linewidth]{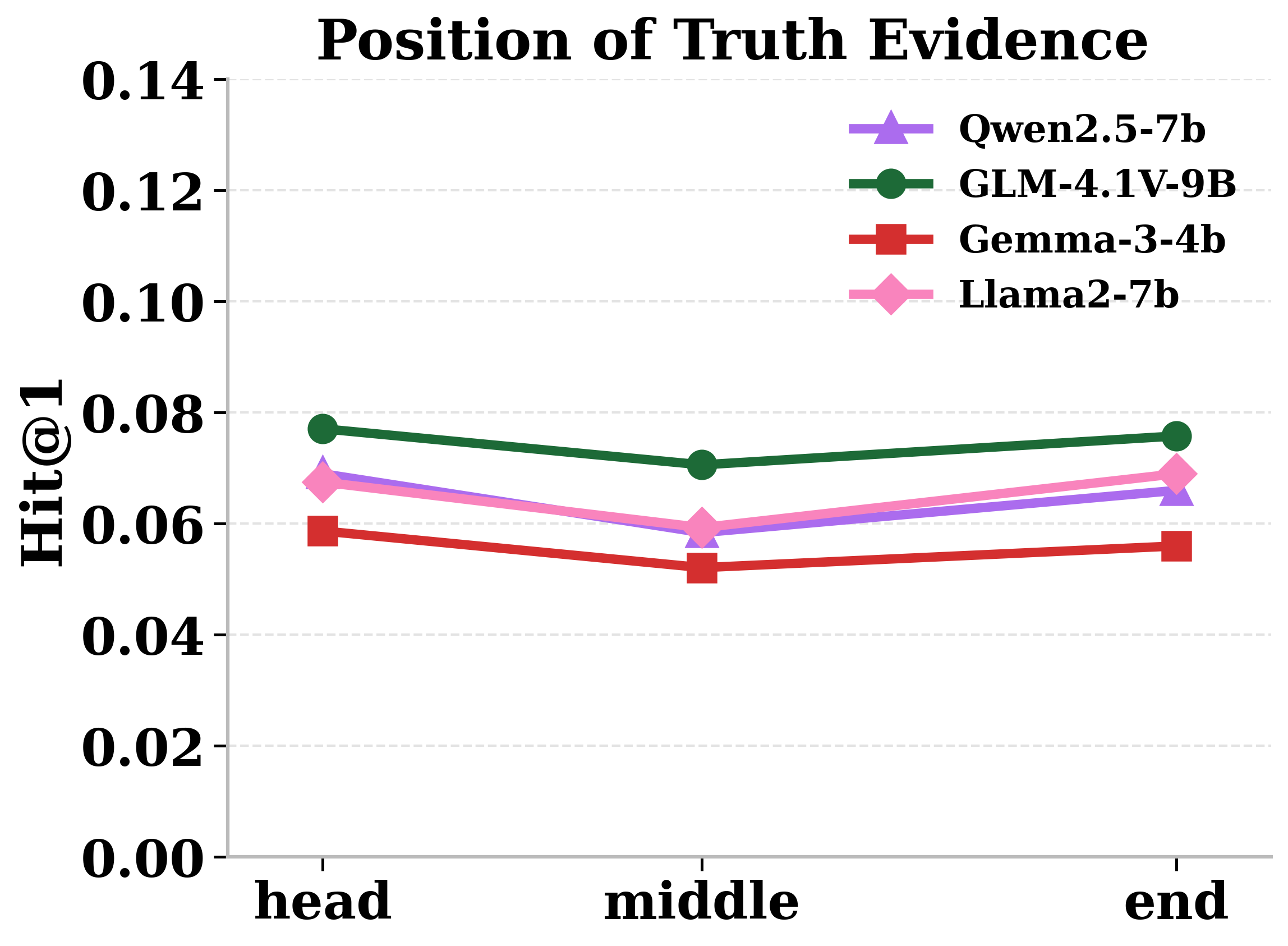}
  \end{minipage}
  \caption{“Lost-in-the-Middle” problem of existing methods. We conduct two position-sensitivity analyses on serialized graph text. Masking the middle segment causes the smallest performance drop, suggesting weak mid-sequence utilization. When ground-truth evidence is placed at the head, middle, or end while other positions are filled with noisy graph information, Hit@1 is lowest for middle placement. This shows that existing LLM/MLLM-based graph-to-text methods underutilize useful evidence in the middle of long contexts.  }
  \label{fig:motivation1}
\end{figure}

\textbf{\textit{Challenge 1: Graph-to-Text Sequences Exacerbate the ``Lost-in-the-Middle'' Problem.}} 
Converting graph structures into text often produces long sequences, making LLMs susceptible to the Lost-in-the-Middle’’ phenomenon \cite{Lost-in-the-Middle,Lost-in-the-Distance,Found-in-the-middle}. To investigate this issue in graph-to-text recommendation, we conduct two position-sensitivity analyses. As shown in Fig.~\ref{fig:motivation1}\footnote{We serialize the unordered user-item graph into a sequence of length $L$ and divide it into three equal segments. In the masking experiment, each segment is replaced with a fixed mask template. In the evidence-placement experiment, ground-truth evidence is placed at different positions alongside two irrelevant graph-text sequences, forming an input of length $3L$. All experiments use the same candidate set and decoding protocol.}, masking the middle segment causes the smallest performance drop, indicating limited utilization of mid-sequence information. Similarly, placing ground-truth evidence in the middle yields the lowest Hit@1 compared with head or end placement. These findings reveal positional bias in graph-to-text recommendation, where important structural information in the middle is underutilized. This raises a key question: can we mitigate positional bias while preserving essential graph structures?

\textbf{\textit{Challenge 2: Sequential Verbalization Causes Semantic Flattening of Graph Structures.}}
In recommendation scenarios, graph structures are heterogeneous and multi-dimensional, involving various types of relations such as co-purchase patterns, preference similarity, and review associations. However, when these graphs are linearized into textual sequences, such as triples, paths, or neighborhood descriptions, their intrinsic multi-hop dependencies and multi-relational topology must be compressed into a one-dimensional sequential form. This transformation often results in semantic flattening \cite{sf3,sf2,sf4}, where rich structural semantics are reduced to surface-level textual patterns.

Consequently, LLMs may struggle to accurately model the underlying graph structure, limiting their ability to effectively leverage structural information for reasoning and recommendation. This issue naturally raises an important question: can we preserve and model graph semantics in a way that avoids the structural information loss caused by sequential verbalization?

To answer the above questions, we propose \textbf{GTRec}. 
Specifically, we first construct a Multimodal Interest Transition Graph (MITG) to alleviate weak interactions in the original user--item graph. By integrating Markov-based collaborative transitions with model-based semantic knowledge, MITG provides a clearer and more compact representation of interest transitions. We then transform MITG into a visual modality, which exposes global topological cues and enables the MLLM to better capture multi-hop and multi-relational patterns, thereby reducing structural information loss caused by graph linearization. However, while visualized graphs preserve global topology, they are limited in expressing fine-grained semantics. Adding extensive textual descriptions could address this limitation but would introduce long sequences, exacerbating the ``Lost-in-the-Middle'' problem. To overcome this, we propose Cluster-based Graph Compression (CGC), which clusters MITG nodes based on semantic similarity and generates composite textual descriptions for each cluster. This preserves the core semantics of interest groups while substantially reducing token consumption. Finally, the visualized MITG and compressed textual semantics are jointly provided to the MLLM as complementary multimodal context, enabling unified structural and semantic reasoning. 
We summarize our contributions as follows:

\begin{itemize}[leftmargin=*]
\item We propose GTRec, a novel MLLM-based recommendation framework that jointly incorporates visualized graph structures and compressed textual semantics for recommendation.
\item We construct the MITG by integrating collaborative and semantic knowledge, and transform it into a visual modality to expose topology-aware structural cues and reduce semantic flattening caused by graph linearization.
\item We propose CGC, which clusters semantically similar nodes and generates concise textual descriptions, reducing input length while preserving core user-interest semantics to mitigate the ``Lost-in-the-Middle'' problem.
\item Experiments on three datasets demonstrate that GTRec consistently outperforms existing methods, validating its effectiveness in improving recommendation performance.
\end{itemize}

\begin{figure*}[htbp]
  \centering
  \includegraphics[width=\textwidth]{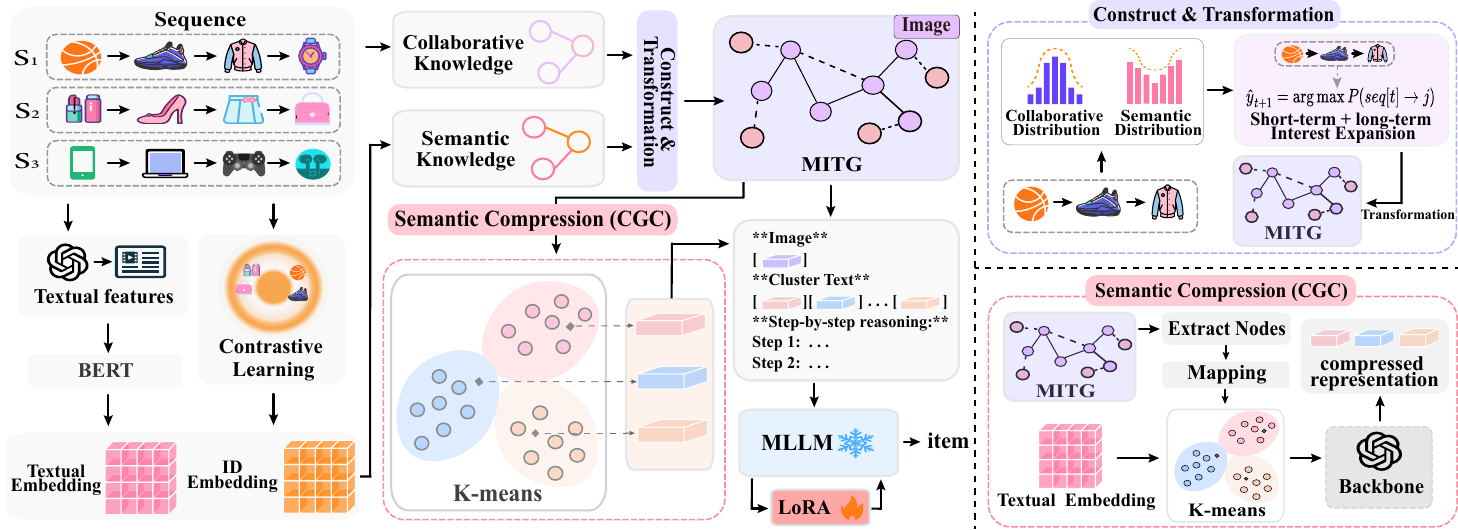}
\caption{GTRec comprises three core modules: 
(1) constructing a Multimodal Interest Transition Graph (MITG) to model collaborative and semantic interest transitions; 
(2) transforming MITG into a visual representation to expose its global topological structure; and 
(3) applying Cluster-based Graph Compression (CGC) to produce compact cluster-level semantic descriptions. 
The resulting visual graph and compressed textual descriptions are then jointly provided to the MLLM as complementary multimodal context for recommendation.}
\label{fig:GTRec}
\end{figure*}

\section{RELATED WORK}
\subsection{LLM-based Recommendation}
With the rapid progress of LLMs, growing research efforts have explored their integration into recommendation systems \cite{QRec,RAMR,MCCRS,GeoCRS,TIGER,EAGER,tmm3,tmm4,tmm5,EyeCRS,DivReason,DisenCRS,MSCRS,ScotRec}. For instance, MoRec \cite{yuan2023morec} explores replacing conventional ID embeddings with language model-derived textual representations, while Recformer \cite{recformer} unifies pre-training and fine-tuning for item text encoding and sequential recommendation. CoLLM \cite{CoLLM} and LLaRA \cite{LLaRA} learn ID embeddings in the semantic space of LLMs to enhance recommendation performance.
Recent studies \cite{LLMREC,GLRec,G-Refer,GaCLLM,kgllmrec} further integrate graph information with LLMs to better capture user preferences. Specifically, G-Refer \cite{G-Refer} retrieves collaborative signals from graphs and converts them into text for semantic alignment with LLMs. GLRec \cite{GLRec} verbalizes heterogeneous relations through meta-path prompts to capture higher-order graph relations, while K-RagRec \cite{kgllmrec} retrieves structured knowledge from knowledge graphs to enrich prompts with entity relations. However, linearizing graph structures into textual sequences often leads to ``Lost-in-the-Middle’’ and semantic flattening. In contrast, we transform graphs into a visual modality to preserve their topology and incorporate compressed textual semantics, enabling joint modeling of global structural relations and fine-grained semantic information.

\subsection{Graph-based Recommendation}
In recommendation scenarios, users and items exhibit associations across multiple dimensions, which naturally give rise to various graph structures, such as user–item interaction graphs  \cite{lightgcn,CORONA}, knowledge graphs \cite{kgllmrec, kgrec1}, and social graphs \cite{social1,social2}. Built upon these structures, many graph-based recommendation methods \cite{lightgcn,Heterogeneous} have been proposed. For instance, classical graph-based methods such as NGCF \cite{NGCF}, LightGCN \cite{lightgcn} leverage information propagation on graphs to capture high-order collaborative signals. Recently, self-supervised learning (SSL) \cite{NCL,SGL,SimGCL,directau,AutoCF} has become a promising paradigm in recommendation systems for providing auxiliary signals and mitigating data sparsity. Representative approaches such as SGL \cite{SGL}, SimGCL \cite{SimGCL}, NCL \cite{NCL} and LightGCL \cite{LightGCL} focus on learning more uniform representations, thereby mitigating the over-smoothing problem inherent in GNNs. Although existing graph-based recommendation methods have achieved notable progress, they still rely primarily on structural signals derived from interaction data while neglecting the complex semantics and reasoning paths beyond the explicit graph topology \cite{kgllmrec,GLRec}. To address this, we propose transforming graph structures into a representation more interpretable by LLMs, enabling semantic reasoning and thereby improving recommendation performance.

\section{PRELIMINARY}

\textbf{Problem Definition.}
Given a user interaction history $H_n=[i_1,i_2,\ldots,i_{n-1}]$, the goal is to predict the next item $i_n$. Conventionally, a recommendation model parameterized by $\theta$ maximizes the conditional probability:
\begin{equation}
\max_{\theta} \log P_{\theta}(i_n\mid H_n).
\end{equation}

In our MLLM-based formulation, we cast next-item prediction as an instruction-following candidate-selection task, where each item is represented by its textual title. For each instance, we construct a candidate set $\mathcal{C}_n=\{i_n\}\cup\mathcal{N}_n$, consisting of the ground-truth item $i_n$ and sampled negatives $\mathcal{N}_n$. The multimodal input comprises the textual titles of historical and candidate items, along with the visualized MITG:
\begin{equation}
x_n=(x_n^{\mathrm{text}},x_n^{\mathrm{image}}).
\end{equation}

Let $y_n$ denote the ground-truth item title. The MLLM is trained to generate $y_n$ autoregressively by minimizing:
\begin{equation}
\mathcal{L}_{\mathrm{MLLM}}
=
-\sum_n\sum_{t=1}^{|y_n|}
\log P_{\theta}
\left(
y_{n,t}\mid y_{n,<t},x_n
\right),
\end{equation}
where $y_{n,t}$ is the $t$-th token of $y_n$. During inference, 
the model selects the candidate with the highest conditional 
likelihood from $\mathcal{C}_n$ as the predicted next item.

\section{METHODOLOGY}
Our approach consists of four main components, as shown in Fig. \ref{fig:GTRec}.
We first construct MITG by integrating Markov-based collaborative transitions with model-based semantic knowledge, and then transform it into a visual modality to preserve topological structures and alleviate semantic flattening.
Next, CGC is introduced to summarize user preferences across multiple interest dimensions.
Finally, a unified multimodal prompt guides the MLLM to generate recommendations by jointly leveraging visual and textual information.

\subsection{Multimodal Knowledge Extraction}
Original user--item graphs often contain noisy interactions, such as accidental clicks and random browsing, which hinder robust interest transition modeling. 
To address this, we integrate collaborative knowledge from item co-occurrence patterns with semantic knowledge from textual and ID modalities to construct more reliable interest transitions. \\
\textbf{Collaborative Knowledge Extraction.}
To capture latent interest transitions, we extract collaborative knowledge from global user behaviors. For each user $u \in \mathcal{U}$, we represent their interaction history as a sequence $\mathcal{S}_u = [i_1, i_2, \dots, i_n]$, where $i_k \in \mathcal{I}$ denotes the $k$-th clicked item. 

We collect directed co-occurrence relations between items across all users to construct a global collaborative graph $\mathcal{G}_\text{collab} = (\mathcal{I}, \mathcal{E}_c)$, where each directed edge $(i_k \rightarrow j) \in \mathcal{E}_c$ indicates that item $j$ follows item $i_k$ in a user's behavior sequence. The edge weight is defined as the normalized transition probability from $i_k$ to $j$ over all users:
\begin{equation}
P_{i_k \rightarrow j}^{(c)} 
= \frac{C_{\,i_k \rightarrow j}}{\sum_{\mathcal{I}} C_{\,i_k \rightarrow j'}} ,
\end{equation}
where $C_{i_k \rightarrow j}$ denotes the number of times item $j$ directly follows item $i_k$ in all user sequences. This probabilistic formulation yields a collaborative transition matrix $\mathbf{P}^{(c)}$, which captures aggregated interest transition patterns at the global level. To suppress noise and enhance the sparsity of the graph structure, we employ a filtering threshold $\tau_c$ and retain only the transition probabilities above the threshold:
\begin{equation}
\mathbf{P}^{(c)}_{i_k \rightarrow j} =
\begin{cases}
P_{i_k \rightarrow j}^{(c)}, & \text{if } P_{i_k \rightarrow j}^{(c)} > \tau_c. \\
0, & \text{otherwise.}
\end{cases}
\end{equation} \\
\textbf{Semantic Knowledge Extraction.}
Semantic knowledge captures latent interest relations beyond behavior sequences by leveraging textual descriptions and ID representations, which model content semantics and behavioral similarities.\\
\textbf{\textit{Textual Semantic Extraction.}} To extract semantic knowledge from the textual modality, we first employ the same backbone model to generate a structured textual description for each item $i$, capturing rich semantic cues:
\begin{equation}
v_i^t = \mathbf{F}_{\text{LLM}}(i; \theta_{prompt}),
\label{eq:5}
\end{equation}
where $\theta_{prompt}$ denotes the prompt template. 
We then feed the generated description $v_i^t$ into a pre-trained RoBERTa encoder~\cite{RoBERTa}, followed by average pooling to obtain a dense semantic representation:
\begin{equation}
e_i^t = \text{AvgPool}\left( \mathbf{F}_{\text{RoBERTa}}(v_i^t; \theta_r) \right),e_i^t \in \mathbb{R}^{d_t},
\end{equation}
where $\theta_r$ represents the parameters of RoBERTa, and $e_i^t$ denotes the textual embedding of item $i$. Collectively, we obtain the full textual embedding set $\mathbf{E}^{\textit{text}} = \{e_{i_1}^\textit{text}, e_{i_2}^\textit{text}, \dots, e_{i_N}^\textit{text}\}$. \\
\textbf{\textit{ID Semantic Extraction.}}
Inspired by \cite{freelunch}, we employ the InfoNCE \cite{infonce} loss to learn semantic representations of user and item IDs. By contrasting positive and negative pairs, the model effectively captures latent semantic similarities among IDs, so that semantically related IDs are closer in the embedding space. We initialize user and item embeddings with two separate embedding tables $\mathbf{E}_u^{id} \in \mathbb{R}^{|\mathcal{U}| \times d}$ and $\mathbf{E}_i^{id} \in \mathbb{R}^{|\mathcal{I}| \times d}$. Given a batch $\mathcal{B}={(u_m, i_m)}_{m=1}^{|\mathcal{B}|}$ of positive user–item pairs,
we normalize the embeddings ${f}(e_u^{id})$ and ${f}(e_i^{id})$.
The ID contrastive loss is defined as the sum of two symmetric InfoNCE objectives:
\begin{align}
\mathcal{L}_{\text{ID}} = - \sum_{m=1}^{|\mathcal{B}|} \Bigg( & 
\log \frac{\exp(f(e^{id}_{u_m})\cdot f(e^{id}_{i_m})/\tau)}{\sum_{n=1}^{|\mathcal{B}|} \exp(f(e^{id}_{u_m})\cdot f(e^{id}_{i_n})/\tau)} \notag\\
& + \log \frac{\exp(f(e^{id}_{i_m})\cdot f(e^{id}_{u_m})/\tau)}{\sum_{n=1}^{|\mathcal{B}|} \exp(f(e^{id}_{i_m})\cdot f(e^{id}_{u_n})/\tau)}
\Bigg),
\end{align}
where $\tau$ is a temperature hyperparameter. Through this symmetric contrastive training, we obtain discriminative and well-distributed ID embeddings $\mathbf{E}_i^{id}$, which can be further utilized to construct ID semantic transition relations among items.

\subsection{MITG Construction}
We construct MITG by integrating Markov-based collaborative transitions with ID-based and text-based semantic preferences. Given a user interaction sequence $\mathcal{S}_u=[i_1,i_2,\ldots,i_n]$, we define the history before position $t$ as $\mathcal{H}_t=[i_1,\ldots,i_{t-1}]$. MITG estimates transitions from historical to candidate items using collaborative knowledge $\mathbf{A}^{(c)}$, textual semantics $\mathbf{E}^{text}$, and ID semantics $\mathbf{E}^{id}$.\\
\textbf{Collaborative Interest Construction.}
We regard each item in $\mathcal{I}$ as a state and the item set $\mathcal{I}$ as the state space. Let $X_t^u$ denote the item state of user $u$ at interaction step $t$, whose observed realization is $i_t$. For collaborative transition modeling, we adopt a first-order Markov assumption:
\begin{equation}
P(X_{t+1}^u=j\mid X_{1:t}^u)
=
P(X_{t+1}^u=j\mid X_t^u).
\end{equation}

Let $C_{i\rightarrow j}$ denote the number of times that item $j$ directly follows item $i$ in the training interaction sequences. The collaborative transition probability is defined as:
\begin{equation}
P^{(c)}(j\mid i)
=
\frac{C_{i\rightarrow j}}
{\sum_{j'\in\mathcal{I}}C_{i\rightarrow j'}}.
\end{equation}
After applying the filtering threshold $\tau_c$, the retained transition
probabilities are renormalized over the remaining successors of each
source item. Therefore, the outgoing transition probabilities form a
normalized distribution.
Given the recent interaction history
$\mathcal{H}_t=[i_{t-k},\ldots,i_{t-1}]$, we use the first-order
transition probabilities to compute the likelihood of a $k$-step interaction path. The collaborative path score of a candidate item $j$
is defined as:
\begin{equation}
Score^{(c)}(j\mid\mathcal{H}_t)
=
\beta_c
\left(
\prod_{\ell=t-k}^{t-2}
P^{(c)}(i_{\ell+1}\mid i_{\ell})
\right)
P^{(c)}(j\mid i_{t-1}),
\end{equation}
where $\beta_c$ is a learnable scaling factor. This quantity is a multi-step path score based on first-order Markov transitions.\\
\textbf{Semantic Interest Construction.}
For the semantic view, we employ two independent GRU networks to encode the ID-based and text-based interaction sequences. The long-term interest representation before position $t$ is defined as:
\begin{equation}
h_\Lambda^{(s)}
=
\mathrm{GRU}^{(s)}
\left(
e^s_{i_1},\ldots,e^s_{i_{t-1}}
\right),
s\in\{id,text\},
\end{equation}
where $h_\Lambda^{(s)}$ is obtained from the last hidden state of the
corresponding GRU. To model short-term interests, we aggregate the most recent $k$
interactions:
\begin{equation}
h_\lambda^{(s)}
=
\sum_{\ell=1}^{k}
\beta_\ell^{(s)}e^s_{i_{t-\ell}},
s\in\{id,text\},
\end{equation}
where $\beta_\ell^{(s)}$ is a learnable weight representing the
importance of the $\ell$-th recent interaction under semantic view $s$.

We then estimate the compatibility between candidate item $j$ and the
user interest under each semantic view:
\begin{equation}
Score^{(s)}(j\mid\mathcal{H}_t)
=
\psi_s
\left(
e_j^s,
h_\Lambda^{(s)},
h_\lambda^{(s)}
\right),
s\in\{id,text\},
\end{equation}
where $\psi_s(\cdot)$ is an MLP-based scoring function.
$Score^{(id)}$ and $Score^{(text)}$ are unnormalized semantic
compatibility scores. These semantic scorers are jointly optimized with
the collaborative scaling factor and fusion parameters using the final
cross-entropy objective defined below.\\
\textbf{Multimodal Interest Fusion.} To integrate preferences from different knowledge sources, we adopt a two-level fusion strategy. We first combine the ID-based and text-based semantic scores using normalized learnable weights:
\begin{equation}
Score^{(sem)}(j\mid\mathcal{H}_t)
=
\sum_{s\in\{id,text\}}
\alpha_s Score^{(s)}(j\mid\mathcal{H}_t),
\end{equation}
where $\alpha_{id}$ and $\alpha_{text}$ are normalized learnable weights for the ID-based and text-based semantic signals.
We then integrate the semantic score with the collaborative path score:
\begin{equation}
\begin{aligned}
Score^{(m)}(j\mid\mathcal{H}_t)
=&\
\alpha_c Score^{(c)}(j\mid\mathcal{H}_t)\\
&+
\alpha_{sem}Score^{(sem)}(j\mid\mathcal{H}_t),
\end{aligned}
\end{equation}
where $\alpha_c$ and $\alpha_{sem}$ are normalized learnable weights that balance the contributions of collaborative and semantic preference signals. We normalize the final multimodal scores into a candidate-conditioned transition distribution:
\begin{equation}
p^{(m)}(j\mid\mathcal{H}_t)
=
\frac{
\exp(Score^{(m)}(j\mid\mathcal{H}_t))
}{
\sum_{v\in\mathcal{I}}
\exp(Score^{(m)}(v\mid\mathcal{H}_t))
}.
\end{equation}
The transition module is trained using the cross-entropy loss:
\begin{equation}
\mathcal{L}_{\mathrm{MITG}}
=
-\sum_{(\mathcal{H}_t,i_t)\in\mathcal{D}_{\mathrm{train}}}
\log p^{(m)}(i_t\mid\mathcal{H}_t).
\end{equation}
During inference, the same distribution is evaluated over the retrieved candidate set $\mathcal{C}_t$ and used to select the Top-$N$ transitions for MITG construction. The parameter $k$ controls the length of the recent interaction path used for short-term interest modeling.\\
\textbf{User-specific MITG Expansion.}
Given the interaction sequence $\mathcal{S}_u$, we expand it into a
user-specific MITG $\mathcal{G}_u=(\mathcal{V}_u,\mathcal{E}_u)$.
For each sequence position $t$, we construct a local candidate pool by
combining collaborative, textual, and ID-based retrieval results:
\begin{equation}
\mathcal{C}_t
=
\mathcal{C}^{(c)}_t
\cup
\mathcal{C}^{(text)}_t
\cup
\mathcal{C}^{(id)}_t,
\end{equation}
where $\mathcal{C}^{(c)}_t$ contains the top-$K$ collaborative
successors of $i_{t-1}$, while $\mathcal{C}^{(text)}_t$ and
$\mathcal{C}^{(id)}_t$ contain the top-$K$ items retrieved according
to their cosine similarities to $h_\lambda^{(text)}$ and
$h_\lambda^{(id)}$, respectively. The candidate pool serves only as a
computationally manageable proposal set during inference. For each candidate $j\in\mathcal{C}_t$, we compute its learned multimodal transition score and construct a scored candidate edge:
\begin{equation}
\mathcal{E}^{cand}_u
=
\left\{
\left(
i_{t-1},
j,
Score^{(m)}(j\mid\mathcal{H}_t)
\right)
\mid
j\in\mathcal{C}_t,\ 1<t\leq n
\right\}.
\end{equation}
The same multimodal score optimized by the cross-entropy objective is
used to rank the retrieved candidates. Since softmax is monotonic,
ranking candidates by $Score^{(m)}$ is equivalent to ranking them by $p^{(m)}$. To obtain a compact graph, we retain the nodes associated with the top-$N$ highest-scoring candidate edges:
\begin{equation}
\mathcal{V}_u
=
\operatorname{TopN}
\left(
\mathcal{E}^{cand}_u;
Score^{(m)}
\right),
\end{equation}
where $\operatorname{Topk}(\cdot)$ returns the nodes associated
with the $N$ candidate edges having the highest multimodal transition scores. The corresponding edge set is defined as:
\begin{equation}
\mathcal{E}_u
=
\left\{
(a,b)
\mid
(a,b,\cdot)\in\mathcal{E}^{cand}_u,\ 
a,b\in\mathcal{V}_u
\right\}.
\end{equation}
The resulting graph $\mathcal{G}_u=(\mathcal{V}_u,\mathcal{E}_u)$
serves as the user-specific MITG, where $N$ is tuned on the validation
set. The Top-$N$ operation is a deterministic graph-sparsification step
rather than probabilistic posterior truncation, and it does not
introduce additional trainable parameters.

\subsection{Visual Transformation of MITG}
User–item interaction graphs involve complex topology and multi-hop transitions. Existing text-based graph linearization methods \cite{GraphGPT,GLRec,G-Refer} often cause semantic flattening and lengthy inputs, limiting LLM reasoning. To address this, we visualize MITG using colors to encode node and edge semantics and spatial layouts to preserve local interactions and global connectivity. This design reduces structural information loss and redundancy while enhancing graph reasoning through MLLMs’ visual and semantic understanding.

Specifically, given the MITG of user $u$ denoted by $\mathcal{G}_u = (\mathcal{V}_u, \mathcal{E}_u)$, we define a graph transformation function $\textit{Transform}(\cdot)$ to transform the graph structure into a visual modality. This transformation process\footnote{We implement it using the Python visualization library \textit{matplotlib} (\url{https://matplotlib.org}), we render the visual MITG using a shell layout to determine node positions.} encodes the structural information of nodes and edges into an image format while preserving their semantic and topological characteristics:
\begin{equation}
\mathcal{G}_u^{visual} = \textit{Transform}(\mathcal{G}_u),
\end{equation}
where $\mathcal{G}_u^{{visual}} \in \mathbb{R}^{H \times W \times 3}$ is the resulting RGB image representation. Nodes of different types are connected using edges with different colors to distinguish their semantic sources (e.g., user–item interaction edges are shown in red, potential interest interaction edges in blue). The spatial layout of nodes preserves both the local interaction order and the global semantic connectivity, facilitating pattern recognition.

\subsection{CGC Construction}
Although the visual MITG effectively preserves graph topology and alleviates semantic flattening, incorporating textual semantics directly would introduce excessively long input sequences, thereby aggravating the “Lost-in-the-Middle” problem. To address this issue, we propose CGC.

In the MITG, user behavior paths often exhibit semantic transitions across different interest dimensions, naturally forming clustered structures. We first perform semantic clustering on the graph structure. Specifically, we apply the $K$-Means \cite{k-means} to the set of all item-level textual embeddings $\mathbf{E}^{text} = \{e_{i_1}^{text}, e_{i_2}^{text}, \dots, e_{i_N}^{text}\}$, resulting in a set of global semantic cluster centers:
\begin{equation}
\mathcal{Z} = \{z_1, z_2, \dots, z_M\},z_m = \frac{1}{|\mathcal{Q}_m|} \sum_{e_i^{text} \in \mathcal{Q}_m} e_i^{text},
\end{equation}
where $\mathcal{Z}$ denotes the set of global semantic cluster centers, $M$ is the total number of clusters, and $\mathcal{Q}_m$ represents the set of node embeddings assigned to the cluster center $z_m$, defined as follows:
\begin{equation}
\mathcal{Q}_m = \left\{ e_i^{text} \;\middle|\; 
\arg\min_{z_j \in \mathcal{Z}} \| e_i^{text} - z_j \|^2 = m \right\}.
\end{equation}

For each user-specific MITG subgraph $\mathcal{G}_u = (\mathcal{V}_u, \mathcal{E}_u)$, we assign each node $v_i\in\mathcal{V}_u$ (with corresponding embedding $e_i^{text}$) to its nearest semantic cluster center as follows:
\begin{equation}
v_i \in \mathcal{O}_f^u 
\;\;\iff\;\; 
f = \arg\min_{m} \| e_i^{text} - z_m \|^2 .
\end{equation}

The user-specific interest graph is then partitioned into:
\begin{equation}
\mathcal{G}_u \Rightarrow \{\mathcal{O}_1^u, \mathcal{O}_2^u, \dots, \mathcal{O}_\mathcal{F}^u\},
\end{equation}
where $\mathcal{O}_f^u$ denotes the $p$-th interest cluster of user $u$, which consists of semantically similar nodes.
$\mathcal{F} \le M$ represents the number of activated clusters for the current user in the global clustering space and reflects the granularity of the user’s interest segmentation.

\begin{figure}[tbp]
    \centering
    \begin{minipage}{0.85\linewidth}
        \centering
        \includegraphics[width=\linewidth]{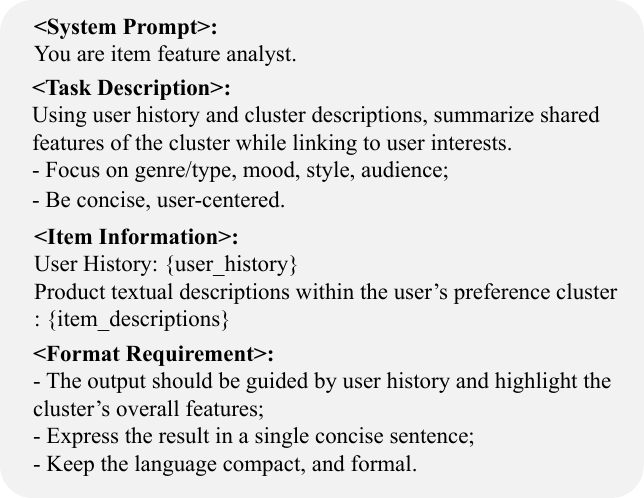}
        \caption{Cluster-based semantics compression template.}
            \label{fig:CGC}
    \end{minipage}
\end{figure}

Next, we generate a summary for each interest cluster based on its constituent nodes and their textual descriptions. These summaries are then used as prompts for the LLM to extract core interest representations. Finally, the multiple interest expressions are fused into a unified prompt:
\begin{equation}
\mathcal{G}_u^{prompt} = \textit{Concat}(\textit{Summary}(\mathcal{O}_1^u), \dots, \textit{Summary}(\mathcal{O}_\mathcal{F}^u)),
\end{equation}
Where $\textit{Summary}(\cdot)$ denotes a prompt template that compresses the node-level descriptions within each interest cluster into representations of the user’s latent preferences, and its specific prompt template is illustrated in Fig. \ref{fig:CGC}. We use the same backbone MLLM (GLM-4.1V-Thinking, Qwen2.5-VL-7B) as in the main pipeline for summary generation.

\begin{figure}[tbp]
    \centering
    \begin{minipage}{\linewidth}
        \centering
        \includegraphics[width=\linewidth]{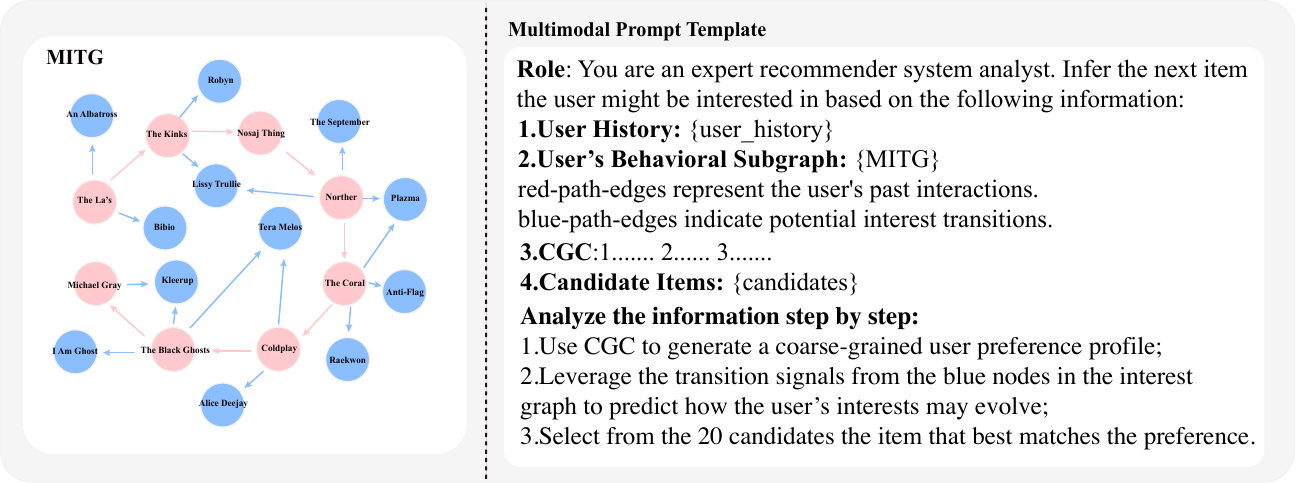}
    \end{minipage}
    \caption{The left shows the MITG: red nodes indicate past interactions and blue nodes indicate potential interest transitions; the right presents the multimodal prompt template, integrating the MITG, user history, the CGC-based preference profile, and candidates to predict the next item of interest.}
    \label{fig:case}
\end{figure}

\subsection{Multimodal Prompt Tuning}

Given the visual representation $\mathcal{G}_u^{visual}$ and textual semantic representation $\mathcal{G}_u^{prompt}$ of MITG, we construct a multimodal prompt to guide the MLLM in generating recommendations. 

\noindent\textbf{Prompt Setting.} The prompt provided to the MLLM mainly consists of four components: the user $u$ historical interaction sequence $\mathcal{S}_u$, the visual representation of the $\mathcal{G}_u^{visual}$, the semantic abstraction of the graph $\mathcal{G}_u^{prompt}$, and the candidate item titles $\{ i_d^{text} \}_{d=1}^{\mathcal{D}}$. Following previous work \cite{LLaRA,vr3}, we fix the candidate set size to $\mathcal{D}$ = 20. Finally, we concatenate all the above information into a unified input prompt with the following structure:
\begin{equation}
\mathcal{X}_u = \textit{Construct}(\mathcal{S}_u,\mathcal{G}_u^{visual},\mathcal{G}_u^{prompt},\{ i_d^{text} \}_{d=1}^{\mathcal{D}},\mathcal{T}_{task}),
\end{equation}
where $\mathcal{T}_{task}$ is a natural language instruction for the recommendation task, the prompt template guides the MLLM to perform step-by-step reasoning over visual and textual inputs, thereby enhancing its capacity to integrate multimodal signals. The prompt template is provided in Fig.~\ref{fig:case}.

\noindent\textbf{Training Objective.} In each instance $\mathcal{X}_u$, the candidate $\{ i_d^{text} \}_{d=1}^{\mathcal{D}}$ contains one ground-truth item $i_*^{text}$ and $\mathcal{D}{-}1$ negative samples. Our objective is to train the model to successfully generate $i_*^{text}$. Specifically, we optimize the following objective function:
\begin{equation}
\min_{\Omega_{\text{LLM}}} \sum -\log P_{\Omega_{\text{LLM}}+\Omega_{\text{vision}}}( i_*^{text} \mid \mathcal{X}_u),
\end{equation}
where we fine-tune only the LLM parameters $\Omega_{\text{LLM}}$ using LoRA, while keeping the vision module $\Omega_{\text{vision}}$ fixed.

During inference, we use the same unified multimodal prompt template $\mathcal{X}_u$ as in training, ensuring consistent input structures and promoting stable recommendation performance.

\section{EXPERIMENTS}

\begin{table}[tbp]
\centering
\small
\caption{Statistics of Datasets. }
\label{tab:data_statis}
\begin{tabular}{lrrr}
\toprule
Dataset& Steam & LastFM & Book-Crossing \\
\midrule
\# Sequence & 13,620 & 10,693 & 2,611 \\ 
\# Item & 1,682 & 3,581 & 4,279 \\ 
\# Interaction & 314,460 & 644,340 & 125,631 \\ 
\bottomrule
\end{tabular}
\end{table}

\begin{table*}[htbp]
\caption{Performance comparison of GTRec with baseline methods. The best results are in bold, and * indicates significant improvements over the strongest baseline ($p$-value $< 0.05$). The strongest baseline is underlined. \textit{Imp.\%} denotes the relative improvement over the corresponding base MLLM; for traditional models, the improvement is not reported.}
\label{tab:overall}
\resizebox{\textwidth}{!}{%
\begin{tabular}{l|ccc|ccc|ccc}
\toprule
\multirow{2}{*}{\textbf{Model}}
& \multicolumn{3}{c|}{\textbf{\textit{Steam}}}
& \multicolumn{3}{c|}{\textbf{\textit{LastFM}}}
& \multicolumn{3}{c}{\textbf{\textit{Book-Crossing}}} \\
& \textbf{\textit{ValidRatio}} & \textbf{\textit{HitRatio@1}} & \textbf{\textit{Imp.\%}}
& \textbf{\textit{ValidRatio}} & \textbf{\textit{HitRatio@1}} & \textbf{\textit{Imp.\%}}
& \textbf{\textit{ValidRatio}} & \textbf{\textit{HitRatio@1}} & \textbf{\textit{Imp.\%}} \\
\midrule
\multicolumn{10}{c}{\textbf{Traditional}} \\
\midrule
GRU4Rec & 1.0000 & 0.3591 & -
        & 1.0000 & 0.2463 & -
        & 1.0000 & 0.3200 & - \\
Caser & 1.0000 & 0.3472 & -
      & 1.0000 & 0.2213 & -
      & 1.0000 & 0.3392 & - \\
SASRec & 1.0000 & 0.3653 & -
       & 1.0000 & 0.2315 & -
       & 1.0000 & 0.3079 & - \\
MITG-Scorer & 1.0000 & 0.3730 & -
            & 1.0000 & 0.2173 & -
            & 1.0000 & 0.3329 & - \\
\midrule
\multicolumn{10}{c}{\textbf{Qwen2.5-VL-7B-Instruct}} \\
\midrule
Qwen2.5-VL & 0.5024 & 0.1035 & 721.20\%
            & 0.5560 & 0.0946 & 695.61\%
            & 0.3462 & 0.0721 & 1567.61\% \\
TALLRec & 0.9934 & 0.3363 & 27.82\%
         & 0.9836 & 0.3194 & 33.20\%
         & 0.9809 & 0.3547 & 19.64\% \\
CoLLM & 0.9883 & 0.3536 & 22.19\%
       & 0.9951 & 0.3585 & 17.30\%
       & 0.9713 & 0.3740 & 14.59\% \\
LLaRA & 0.9912 & 0.3644 & 18.22\%
       & 0.9959 & 0.3607 & 16.49\%
       & 0.9856 & 0.3792 & 11.37\% \\
GraphGPT & 0.9890 & 0.3484 & 23.93\%
          & 0.9942 & 0.3393 & 24.05\%
          & 0.9713 & 0.3980 & 7.68\% \\
GLRec & 0.9890 & \underline{0.3942} & 9.53\%
       & 0.9883 & \underline{0.3840} & 10.27\%
       & 0.9809 & \underline{0.4027} & 5.38\% \\
G-Refer & 0.9941 & 0.3781 & 13.61\%
         & 0.9961 & 0.3717 & 13.02\%
         & 0.9713 & 0.3816 & 12.30\% \\
\textbf{GTRec} & 0.9956 & \textbf{0.4289*} & -
               & 0.9940 & \textbf{0.4210*} & -
               & 0.9569 & \textbf{0.4350*} & - \\
\midrule
\multicolumn{10}{c}{\textbf{GLM-4.1V-9B-Thinking}} \\
\midrule
GLM-4.1V & 0.5841 & 0.1455 & 408.87\%
          & 0.6460 & 0.1558 & 351.08\%
          & 0.4012 & 0.1221 & 766.03\% \\
TALLRec & 0.9934 & 0.3480 & 25.10\%
         & 0.9836 & 0.3689 & 25.12\%
         & 0.9713 & 0.3425 & 27.53\% \\
CoLLM & 0.9896 & 0.3662 & 19.34\%
       & 0.9934 & 0.3912 & 16.82\%
       & 0.9809 & 0.3710 & 16.58\% \\
LLaRA & 0.9975 & 0.3814 & 13.67\%
       & 0.9960 & 0.4070 & 12.00\%
       & 0.9856 & 0.3742 & 15.03\% \\
GraphGPT & 0.9875 & 0.3645 & 20.15\%
          & 0.9974 & 0.3865 & 17.77\%
          & 0.9713 & 0.3807 & 14.73\% \\
GLRec & 0.9912 & \underline{0.4019} & 8.56\%
       & 0.9960 & \underline{0.4203} & 8.45\%
       & 0.9665 & \underline{0.4069} & 7.88\% \\
G-Refer & 0.9934 & 0.3862 & 12.72\%
         & 0.9891 & 0.4150 & 10.60\%
         & 0.9617 & 0.3921 & 12.51\% \\
\textbf{GTRec} & 0.9919 & \textbf{0.4360*} & -
               & 0.9954 & \textbf{0.4561*} & -
               & 0.9809 & \textbf{0.4325*} & - \\
\bottomrule
\end{tabular}%
}
\end{table*}

\subsection{Experimental Setup} 
\subsubsection {Datasets} 
We evaluate our framework on three widely used benchmark datasets: \textit{Steam} \cite{Steam}, \textit{LastFM} \cite{LastFM}, and \textit{Book-Crossing}. 
For all three datasets, we organize each user’s interaction sequences in chronological order and split the data into train, validation, and test sets with a ratio of 8:1:1. This partitioning ensures that no future interactions appear in the training set, thereby avoiding potential information leakage. 
Detailed statistics for each dataset are provided in Table \ref{tab:data_statis}.

\subsubsection {Baselines} To evaluate the performance of our proposed model, we compare it with the following baseline methods:
\begin{enumerate}[leftmargin=*] 
    \item Traditional Methods: 
    \begin{itemize}[leftmargin=0pt, align=left]
        \item \textbf{GRU4Rec} \cite{GRU4Rec} is a GRU-based sequential method that predicts the next item from user behavior sequences.
        \item \textbf{Caser} \cite{Caser} captures sequential patterns by applying convolutional operations on user interaction sequences.
        \item \textbf{SASRec} \cite{SASRec} builds on self-attention to model long-range dependencies in user interaction sequences. 
        \item \textbf{MITG-Scorer} performs recommendation using GTRec’s trained multimodal interest transition scorer.
    \end{itemize}
\end{enumerate}
\begin{enumerate}[leftmargin=*, resume] 
    \item LLMs-based Methods: 
    \begin{itemize}[leftmargin=0pt, align=left]
        \item \textbf{Qwen2.5-VL} is Alibaba’s 7B-parameter MLLM supporting image, text, and video inputs.
        \item \textbf{GLM-4.1V} is a 9B parameter MLLM proposed by Zhipu AI, which incorporates chain-of-thought prompting and RLCS to excel in multimodal reasoning.
        \item \textbf{TALLRec\footnote{\label{myfootnote} TALLRec and CoLLM are originally designed for point-wise binary (YES/NO) prediction, while GLRec supports both point-wise and pair-wise instructions. We adapt all models to our setting by selecting the next item from a given candidate set.}} \cite{TALLRec} efficiently adapts LLMs to recommendation tasks using Alpaca instruction tuning and LoRA, achieving strong performance.
        \item \textbf{CoLLM\footnotemark[3]}\cite{CoLLM} integrates collaborative embeddings into the token space of LLM, enabling collaboration–text fusion without altering the LLM architecture.
        \item \textbf{LLaRA} \cite{LLaRA} aligns ID-based embeddings from traditional recommenders with LLM input space via hybrid prompts and curriculum tuning, enabling joint optimization of user behavior modeling and response generation.
    \end{itemize}
\end{enumerate}
\begin{enumerate}[leftmargin=*, resume] 
    \item Methods Combining Graphs and LLMs: 
    \begin{itemize}[leftmargin=0pt, align=left]
        \item \textbf{GraphGPT\footnote{\label{myfootnote} 
        GraphGPT is originally designed for general graph tasks, and we adapt it for user modeling in recommendation systems. G-Refer is originally designed for explainability in recommendations, and we adapt it for prediction.}}\cite{GraphGPT} adopts instruction tuning to align textual and graph-structured data for complex understanding.
        \item \textbf{GLRec\footnotemark[3]}\cite{GLRec} introduces a meta-path prompt constructor to convert graph interactions into language prompts for personalized recommendation.
        \item \textbf{G-Refer\footnotemark[4]}\cite{G-Refer} introduces a graph retrieval-augmented LLM that extracts structural and semantic CF signals via hybrid retrieval for explainable recommendation.
    \end{itemize}
\end{enumerate}

\begin{table}[tbp]
\centering
\caption{Ablation study results on \textit{Steam}, \textit{LastFM}, and \textit{Book-Crossing} with \textit{HitRatio@1} under the Qwen2.5-VL-7B-Instruct backbone.}
\setlength{\tabcolsep}{2pt}
\begin{tabular}{lccc}
\toprule
\textbf{Ablations} & \textit{\textbf{Steam}} & \textit{\textbf{LastFM}} & \textit{\textbf{Book-Crossing}} \\
\midrule
GTRec & 0.4289 & 0.4210 & 0.4350 \\
w/o MITG   & 0.3648 ($\downarrow 14.93\%$) & 0.3529 ($\downarrow 16.18\%$) & 0.3710 ($\downarrow 14.71\%$) \\
w/o CGC    & 0.4016 ($\downarrow 6.37\%$)  & 0.3924 ($\downarrow 6.80\%$)  & 0.4087 ($\downarrow 6.03\%$) \\
w/o S-ID   & 0.3928 ($\downarrow 8.42\%$)  & 0.3955 ($\downarrow 6.05\%$)  & 0.4015 ($\downarrow 7.70\%$) \\
w/o S-Text & 0.3970 ($\downarrow 7.44\%$)  & 0.3935 ($\downarrow 6.55\%$)  & 0.4172 ($\downarrow 4.10\%$) \\
w/o C      & 0.4158 ($\downarrow 3.05\%$)  & 0.4074 ($\downarrow 3.22\%$)  & 0.4120 ($\downarrow 5.29\%$) \\
\bottomrule
\end{tabular}
\label{tab:ablation-qwen}
\end{table}

\begin{table}[tbp]
\centering
\caption{Ablation study results on \textit{Steam}, \textit{LastFM}, and \textit{Book-Crossing} with \textit{HitRatio@1} under the GLM-4.1V-9B-Thinking backbone.}
\setlength{\tabcolsep}{2pt}
\begin{tabular}{lccc}
\toprule
\textbf{Ablations} & \textit{\textbf{Steam}} & \textit{\textbf{LastFM}} & \textit{\textbf{Book-Crossing}} \\
\midrule
GTRec & 0.4360 & 0.4561 & 0.4325 \\
w/o MITG   & 0.3720 ($\downarrow 14.68\%$) & 0.3861 ($\downarrow 15.34\%$) & 0.3759 ($\downarrow 13.08\%$) \\
w/o CGC    & 0.4107 ($\downarrow 5.79\%$)  & 0.4260 ($\downarrow 6.59\%$)  & 0.4125 ($\downarrow 4.62\%$) \\
w/o S-ID   & 0.3992 ($\downarrow 8.44\%$)  & 0.4284 ($\downarrow 6.08\%$)  & 0.3950 ($\downarrow 8.68\%$) \\
w/o S-Text & 0.4081 ($\downarrow 6.40\%$)  & 0.4310 ($\downarrow 5.49\%$)  & 0.4027 ($\downarrow 6.90\%$) \\
w/o C      & 0.4207 ($\downarrow 3.51\%$)  & 0.4380 ($\downarrow 3.97\%$)  & 0.4180 ($\downarrow 3.34\%$) \\
\bottomrule
\end{tabular}
\label{tab:ablation-glm}
\end{table}

\subsubsection {Baseline Adaptation and Fairness Protocol}
To ensure a fair comparison, all methods use identical chronological data splits, user interaction histories, ground-truth next items, and candidate sets. Candidate items are represented by textual titles, and all LLM-based methods are instructed to select or generate the next-item title from the candidate set. The common prompt includes user histories and candidate titles, while graph-based methods additionally encode graph contexts from the same user–item interaction data using their respective mechanisms. 
We adapt TALLRec and CoLLM from their original objectives to the next-item selection task. GraphGPT, GLRec, and G-Refer retain their respective graph-processing mechanisms, using serialized graph inputs, graph-based prompts, and graph retrieval contexts, respectively. All LLM-based methods use frozen backbones with identical LoRA configurations and backbone-specific training budgets, including the optimizer, learning rate, batch size, training epochs, and random seed.


\subsubsection{Evaluation Metrics}
Following previous work \cite{LLaRA, vr2, vr3}, we adopt \textit{HitRatio@1} as the primary metric, measuring whether the predicted item matches the ground-truth target. For each instance, we construct a candidate set with one positive item and 19 negative items uniformly sampled from the user’s uninteracted items. All methods share the same candidate sets for fair comparison. For LLM-based methods, we additionally report \textit{ValidRatio}, measuring the proportion of generated outputs that fall within the candidate set.
To further assess ranking quality, we evaluate GTRec in a retrieve-and-rerank setting in Section~\ref{sec:topk-ranking}, where SASRec retrieves the top-100 candidates for GTRec to rerank. We report \textit{HitRatio@5}, \textit{HitRatio@10}, \textit{NDCG@5}, and \textit{NDCG@10}. \textit{HitRatio@$K$} measures whether the ground-truth item appears in the top-$K$ results, while \textit{NDCG@$K$} additionally considers its ranking position.

\subsubsection{Implementation Details}
Experiments are conducted on 8 NVIDIA A6000 GPUs. All LLM-based methods use identical data splits, user histories, candidate sets, and backbone-specific training budgets. We adopt Qwen2.5-VL-7B-Instruct \cite{Qwen2.5-VL-7B} and GLM-4.1V-9B-Thinking \cite{GLM-4.1V-Thinking} as backbones, trained for 10 epochs with LoRA ($r=8$, $\alpha=16$, dropout $0.05$) and AdamW (learning rate $2\times10^{-4}$), without scheduling or warm-up. We use a per-device batch size of 1, gradient accumulation of 4, float16 precision, and seed 2026.
All MITG interaction statistics are computed from training data only. The MITG scorer is trained separately for 100 epochs using AdamW (learning rate $10^{-3}$, weight decay $10^{-4}$, batch size 512). Its context encoder uses a single-layer unidirectional GRU (hidden size 128) for long-term interests, while the latest $k$ interactions capture short-term interests. Textual and ID scorers use two-layer MLPs (hidden size 128), and the modality-weighting MLP has a hidden size of 64. All modalities have an initial temperature of 1.0. We set retrieval size $K=300$ and image resolution to $910\times910$, tuning node budget $N\in{20,30,40,50}$, transition step $k\in{1,\ldots,5}$, and active CGC clusters $\mathcal{F}\in{1,\ldots,5}$. Traditional recommenders use Adam~\cite{adam} (learning rate $0.001$, embedding size 64, batch size 256).

\begin{table*}[tbp]
\centering
\renewcommand{\arraystretch}{1.15}
\caption{Performance comparison of various LLM-based recommenders as
candidate-item rerankers for Top-$k$ recommendation. The best and
second-best results are highlighted in bold and underlined, respectively.
An asterisk indicates a statistically significant improvement
($p$-value $<0.05$).}
\label{tbl:topktable}
\begin{tabular*}{\textwidth}{@{\extracolsep{\fill}}ll|cccc|cccc@{}}
\toprule
\multirow{2}{*}{\textbf{Backbone}}
& \multirow{2}{*}{\textbf{Method}}
& \multicolumn{4}{c}{\textbf{\textit{Steam}}}
& \multicolumn{4}{c}{\textbf{\textit{LastFM}}} \\[2pt]
\cmidrule(lr){3-6}
\cmidrule(lr){7-10}
&
& \textbf{HitRatio@5}
& \textbf{HitRatio@10}
& \textbf{NDCG@5}
& \textbf{NDCG@10}
& \textbf{HitRatio@5}
& \textbf{HitRatio@10}
& \textbf{NDCG@5}
& \textbf{NDCG@10} \\
\midrule

\multirow{4}{*}{Qwen2.5-VL}
& SASRec
& 0.0634 & 0.1037 & 0.0328 & 0.0469
& 0.0572 & 0.0861 & 0.0342 & 0.0435 \\

& CoLLM
& 0.0587 & 0.0973 & 0.0302 & 0.0437
& 0.0592 & 0.0912 & 0.0354 & 0.0462 \\

& GLRec
& \underline{0.0656}
& \underline{0.1075}
& \underline{0.0349}
& \underline{0.0484}
& \underline{0.0618}
& \underline{0.0988}
& \underline{0.0373}
& \underline{0.0495} \\

& GTRec
& \textbf{0.0677}*
& \textbf{0.1115}*
& \textbf{0.0360}*
& \textbf{0.0502}
& \textbf{0.0633}*
& \textbf{0.1017}*
& \textbf{0.0385}*
& \textbf{0.0509} \\

\midrule

\multirow{4}{*}{GLM-4.1V}
& SASRec
& 0.0634 & 0.1037 & 0.0328 & 0.0469
& 0.0572 & 0.0861 & 0.0342 & 0.0435 \\

& CoLLM
& 0.0624 & 0.1016 & 0.0315 & 0.0455
& 0.0610 & 0.0934 & 0.0362 & 0.0474 \\

& GLRec
& \underline{0.0679}
& \underline{0.1112}
& \underline{0.0353}
& \underline{0.0492}
& \underline{0.0621}
& \underline{0.0977}
& \underline{0.0378}
& \underline{0.0506} \\

& GTRec
& \textbf{0.0710}*
& \textbf{0.1172}*
& \textbf{0.0370}*
& \textbf{0.0519}
& \textbf{0.0642}*
& \textbf{0.1034}*
& \textbf{0.0393}*
& \textbf{0.0520} \\

\bottomrule
\end{tabular*}
\end{table*}

\subsection{Overall Performance}

In this section, we compare GTRec with various baseline models across two mainstream MLLM backbones and three benchmark datasets to answer RQ1. As shown in Table~\ref{tab:overall}, GTRec consistently outperforms the strongest baseline in terms of \textit{HitRatio@1} across all three datasets and both MLLM backbones, demonstrating its effectiveness and generalizability across different recommendation scenarios.

Traditional sequential methods generally perform worse than GTRec, as they rely primarily on interaction sequences without explicitly modeling rich semantic and structural information. Directly applying MLLMs yields low \textit{ValidRatio} and \textit{HitRatio@1}, indicating that zero-shot reasoning alone is insufficient for recommendation. Although LLM-based methods improve performance by incorporating textual and sequential features, their limited structural modeling capabilities constrain their effectiveness. Graph-enhanced methods further improve performance by leveraging graph information. However, MITG-Scorer, which directly uses the learned multimodal interest transition scorer without involving MLLM reasoning, still performs worse than GTRec. This suggests that GTRec’s gains arise not only from MITG construction but also from multimodal reasoning over visualized graph topology and compressed semantic descriptions. Together, these components enable GTRec to achieve superior recommendation performance across different backbones and datasets.

\subsection{Ablation Study}
To systematically investigate the contribution of each component in GTRec to answer RQ2, we design a series of ablation models:
(1) \textbf{w/o MITG}: we remove the visual-modality input of MITG to assess its contribution.
(2) \textbf{w/o CGC}: we exclude the CGC module, forcing the model to rely solely on image representations without compressing textual semantics.
(3) \textbf{w/o S-ID}: we remove ID-based semantics when constructing MITG.
(4) \textbf{w/o S-Text}: we remove text semantics when constructing MITG.
(5) \textbf{w/o C}: we remove collaborative knowledge when constructing MITG. As shown in Tables \ref{tab:ablation-qwen} and \ref{tab:ablation-glm}, removing visual MITG causes the largest performance drop, highlighting its importance in capturing user interests. Excluding CGC also degrades performance, demonstrating the value of graph semantic clustering in refining textual representations and enhancing graph modeling. Similarly, removing ID-based semantics, textual semantics, or collaborative knowledge leads to performance declines, confirming their complementary contributions. These trends remain consistent across Qwen2.5-VL and GLM-4.1V, validating the effectiveness of each GTRec component.

\subsection{Textual vs. Visual MITG}
\begin{figure}[tbp]
    \centering
    \begin{minipage}{.49\linewidth}
        \centering
        \includegraphics[width=\linewidth]{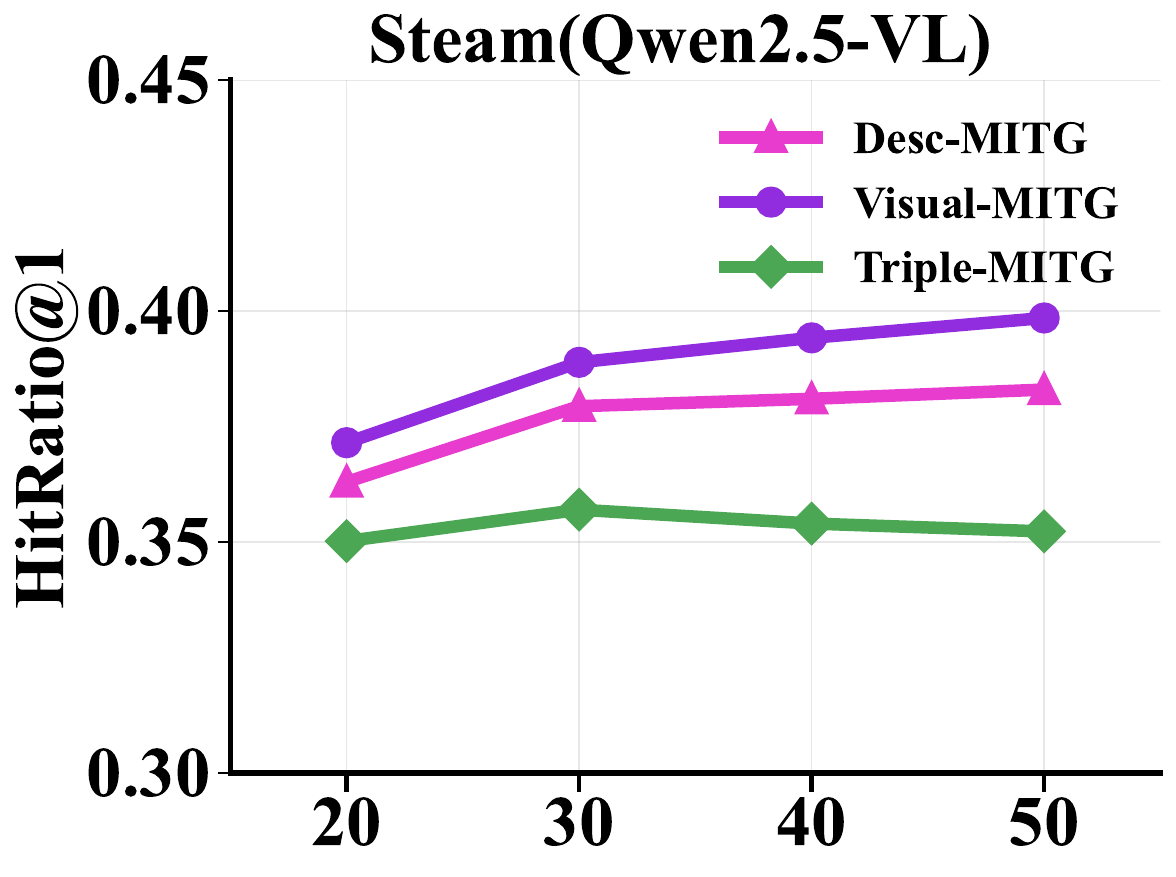}
    \end{minipage}\hfill
    \begin{minipage}{.49\linewidth}
        \centering
        \includegraphics[width=\linewidth]{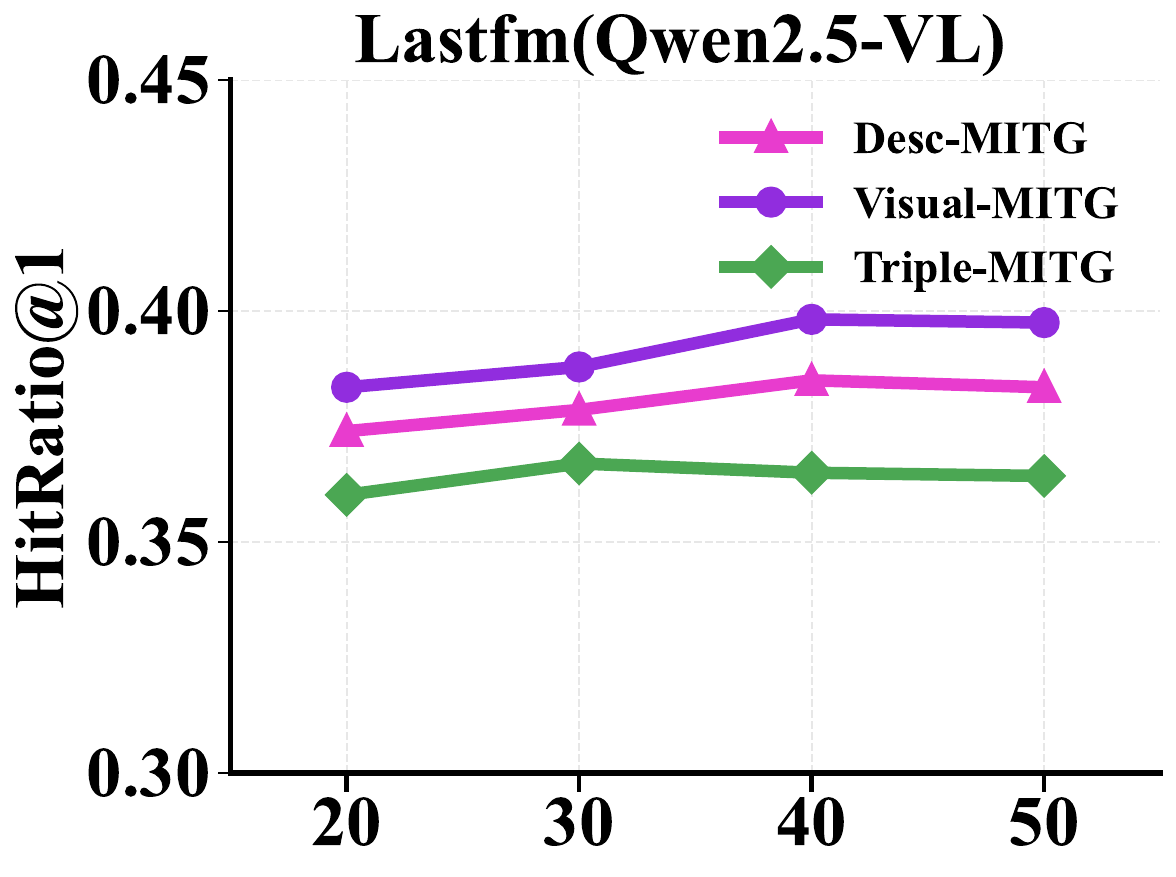}
    \end{minipage}

    \begin{minipage}{0.235\textwidth}
        \centering
        \includegraphics[width=\textwidth]{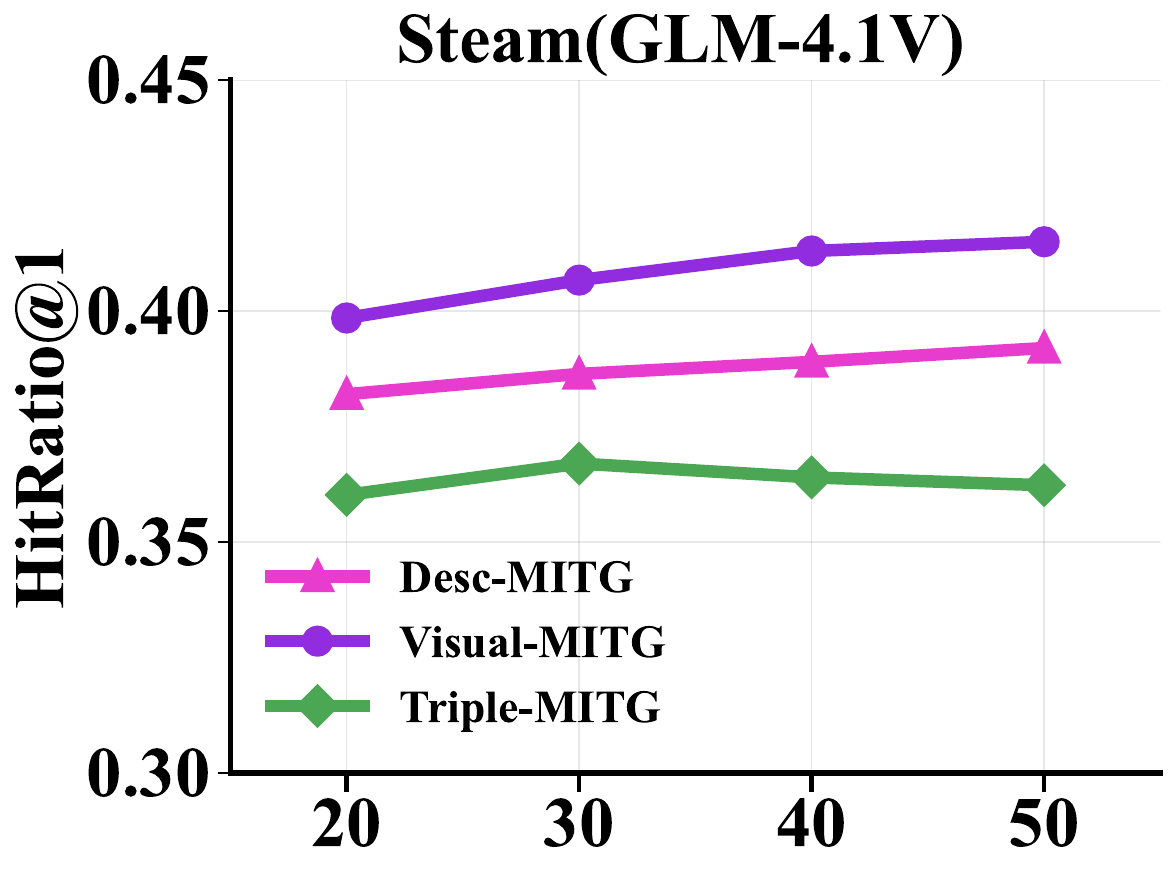}
    \end{minipage}\hfill
    \begin{minipage}{0.235\textwidth}
        \centering
        \includegraphics[width=\textwidth]{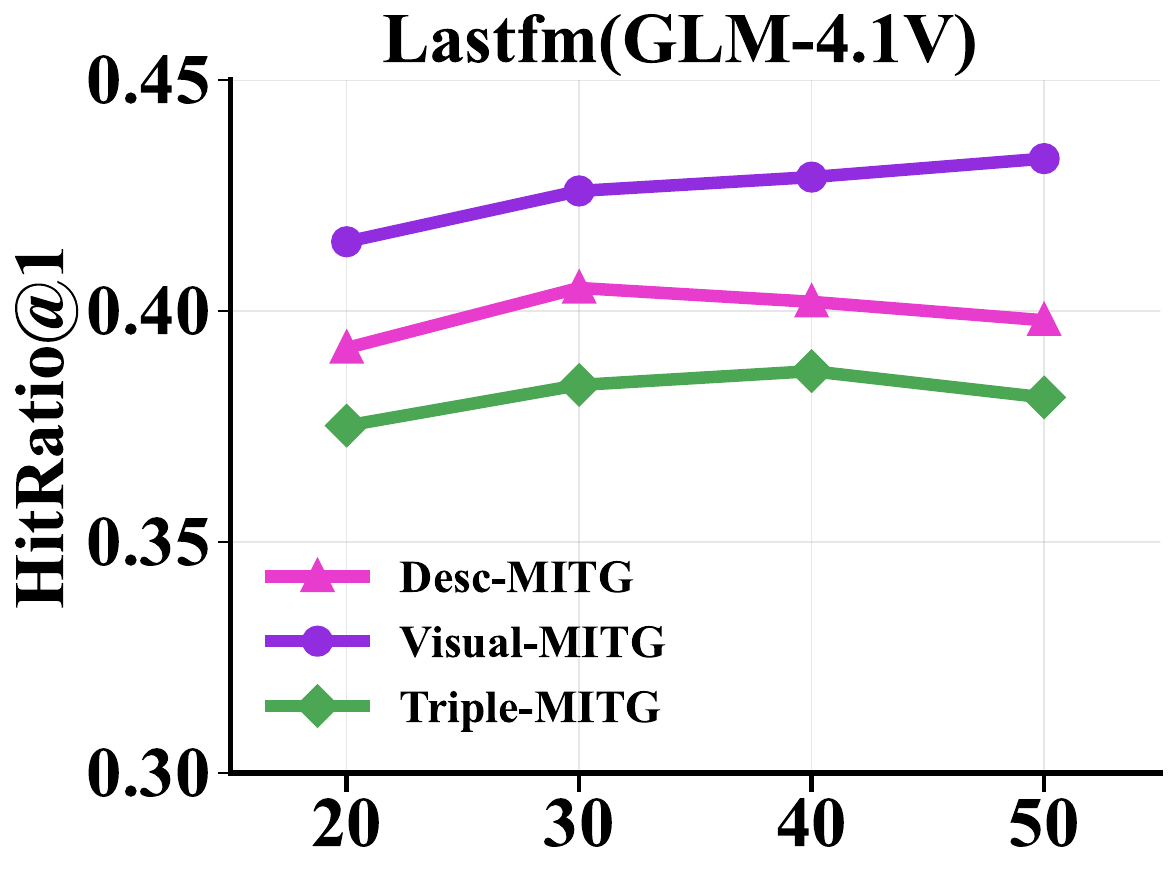}
    \end{minipage}

    \caption{Comparison of three MITG formats across different datasets (Steam and Lastfm) and MLLMs (Qwen2.5-VL and GLM-4.1V).}
    \label{fig:combined}
\end{figure}

\begin{figure}[tbp]
    \centering
    \begin{minipage}{.49\linewidth}
        \centering
        \includegraphics[width=\linewidth]{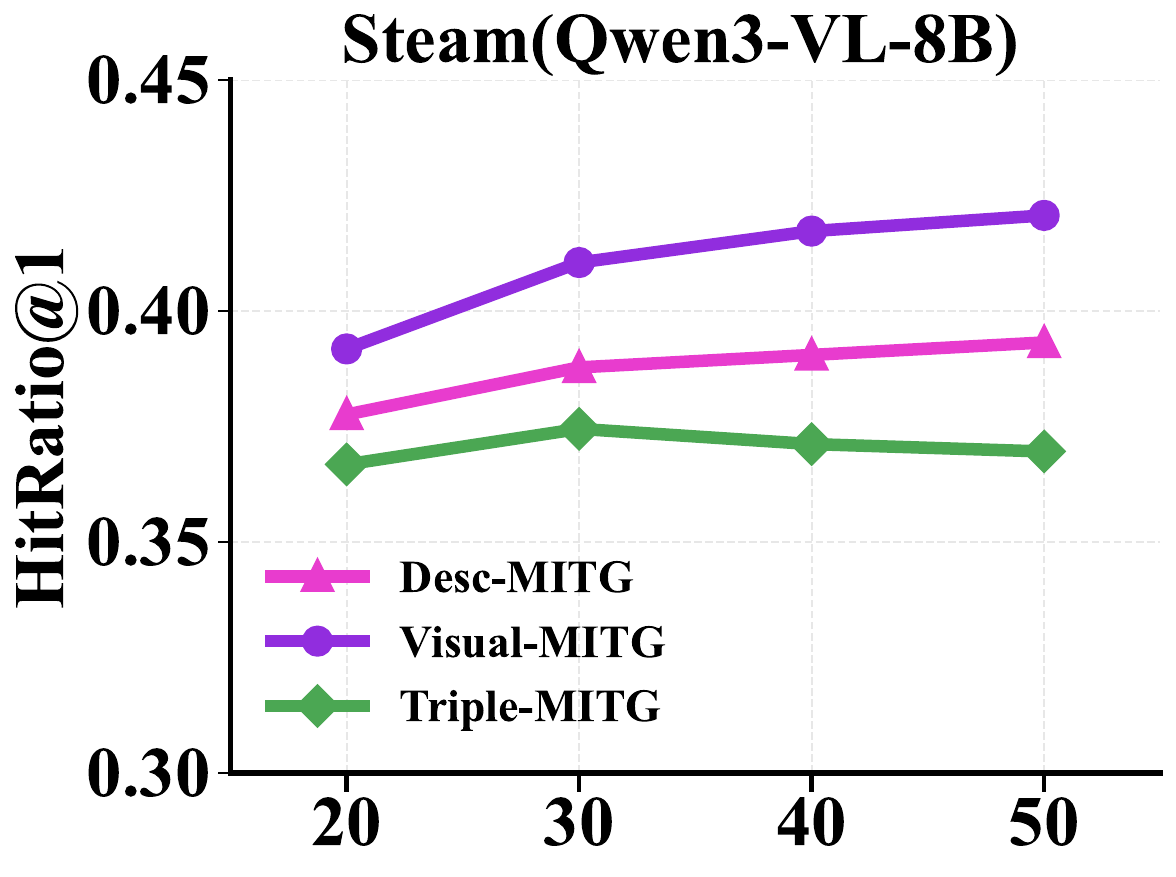}
    \end{minipage}\hfill
    \begin{minipage}{.49\linewidth}
        \centering
        \includegraphics[width=\linewidth]{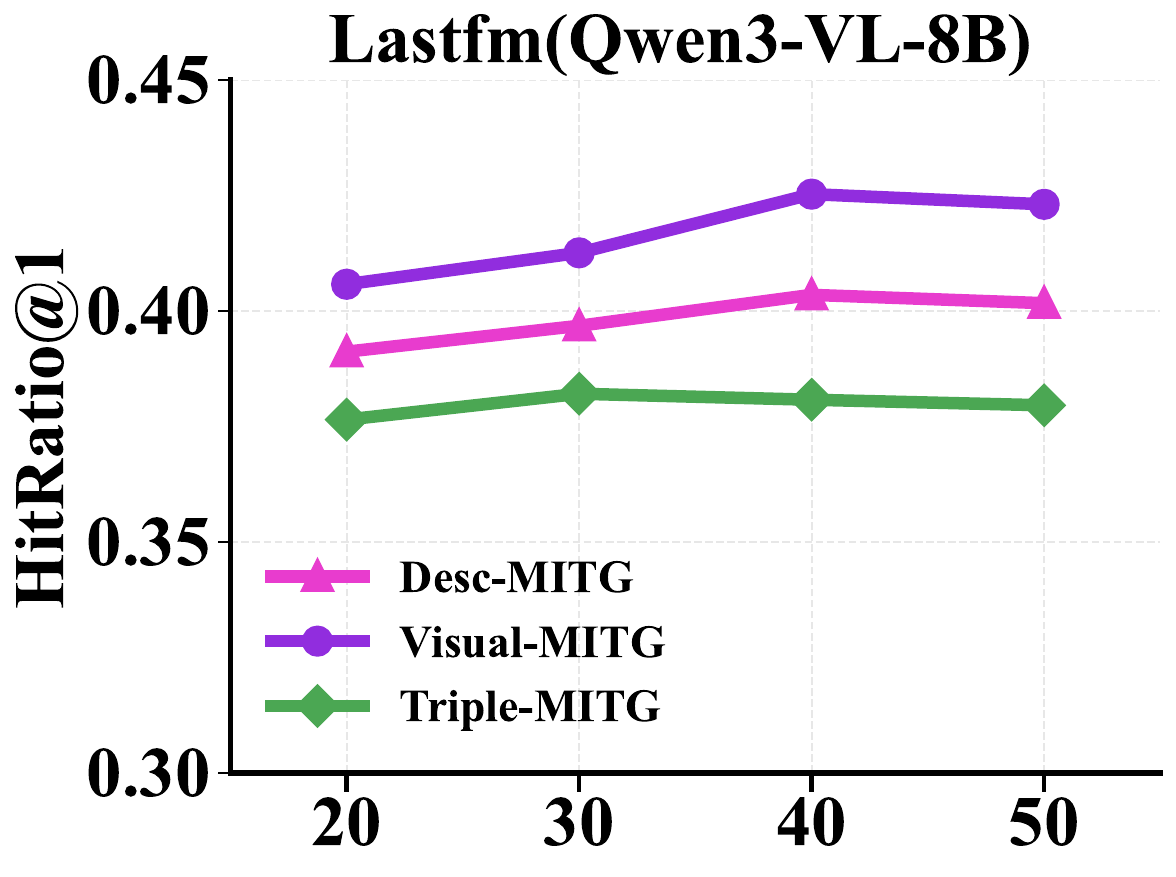}
    \end{minipage}
    \begin{minipage}{0.235\textwidth}
        \centering
        \includegraphics[width=\textwidth]{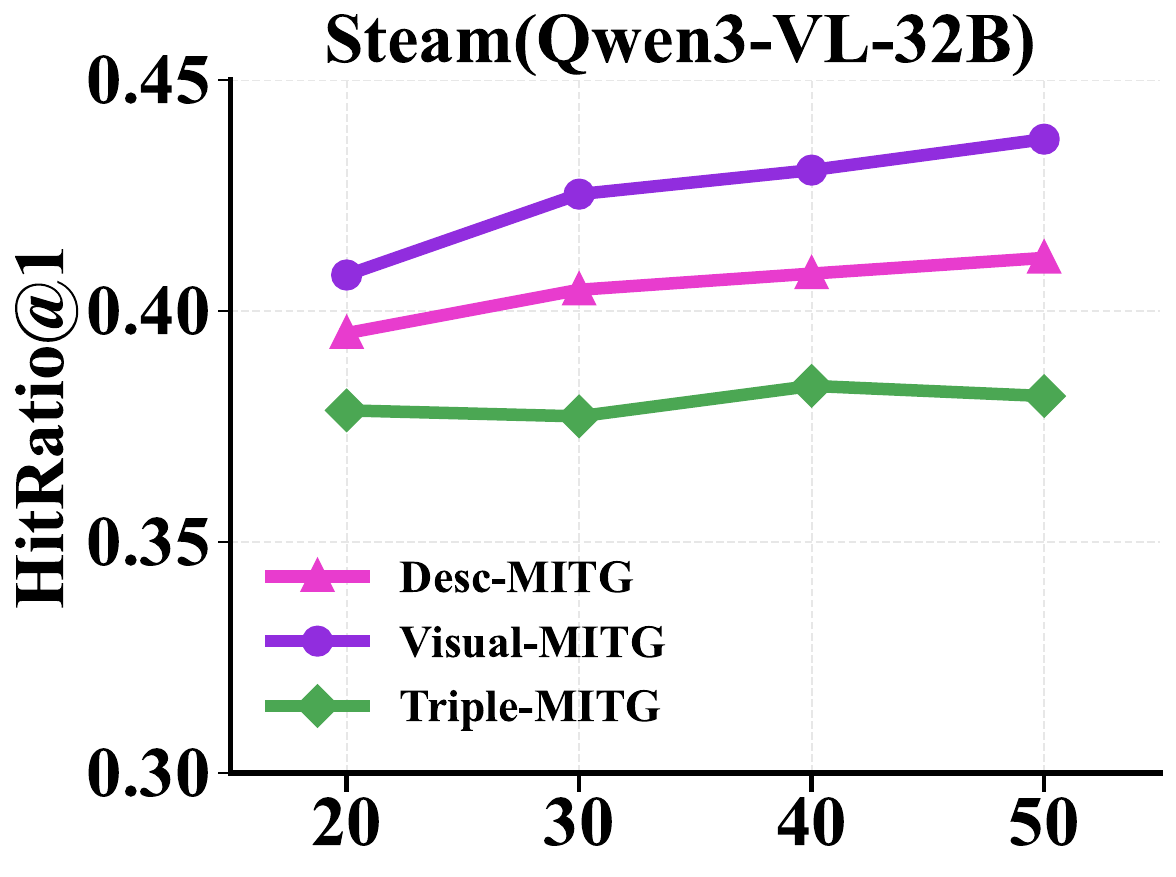}
    \end{minipage}\hfill
    \begin{minipage}{0.235\textwidth}
        \centering
        \includegraphics[width=\textwidth]{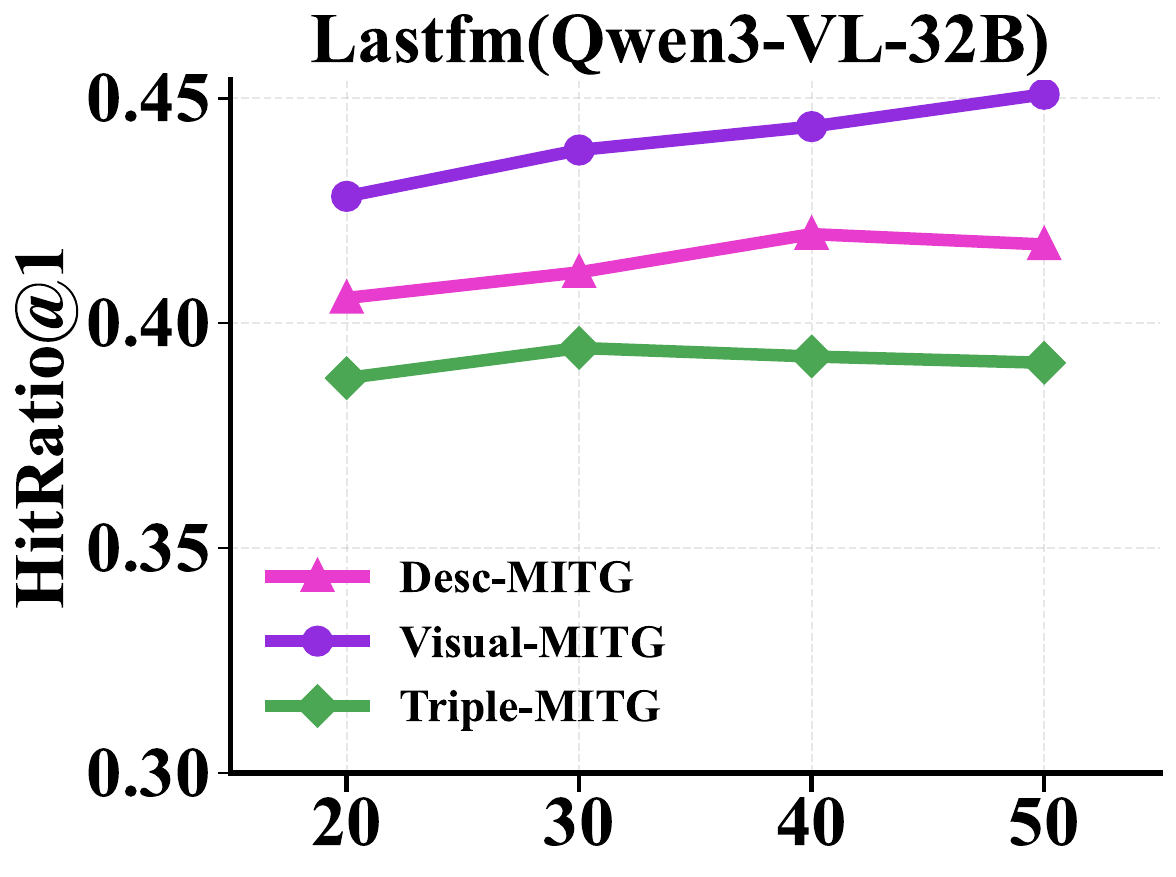}
    \end{minipage}
    \caption{HitRatio@1 of the three MITG formats across Steam and LastFM using larger-scale Qwen3-VL models (8B and 32B).}
    \label{fig:combined2}
\end{figure}

In RQ3, we compare three different representations of MITG: \textbf{Triple-MITG} (graph relations converted into triple), \textbf{Desc-MITG} (sequentialized descriptions based on node text), and \textbf{Visual-MITG} (transforming the graph into visual modality). As shown in Fig.~\ref{fig:combined}, across both the Steam and Lastfm datasets, Visual-MITG consistently achieves the best performance in terms of \textit{HitRatio@1} on two MLLMs (Qwen2.5-VL and GLM-4.1V). In contrast, Triple-MITG performs the worst due to severe semantic flattening, while Desc-MITG shows moderate improvements but still suffers from long-sequence limitations. Overall, these results demonstrate that transforming graphs into a visual modality better preserves structural integrity and multi-relational dependencies, thereby alleviating semantic flattening in graph-to-text recommendation.

To examine whether these advantages generalize to stronger vision-language backbones, we extend our evaluation to Qwen3-VL-8B and Qwen3-VL-32B \cite{qwen3-VL}. As shown in Fig. 6, Visual-MITG consistently achieves the highest HitRatio@1 across both datasets and all candidate-set sizes, outperforming Desc-MITG and Triple-MITG. Compared with Qwen2.5-VL-7B, the performance gap between Visual-MITG and Desc-MITG slightly widens on Qwen3-VL backbones, while remaining comparable to that on GLM-4.1V-9B. These results confirm the robustness and generalizability of Visual-MITG across different backbone architectures and scales.

\subsection{Top-K Ranking Evaluation}
\label{sec:topk-ranking}
Beyond generation-based evaluation, we further assess GTRec in a retrieve-and-rerank setting. For each user, SASRec~\cite{SASRec} retrieves a fixed Top-$100$ candidate set shared by all methods. Each reranking method independently evaluates candidates by estimating their likelihood of being the user’s next interaction. For LLM-based methods, relevance scores are derived from normalized output probabilities, while SASRec retains its original prediction scores without reranking.
GTRec jointly takes the candidate item title, visual MITG, and compressed textual representation as input to estimate candidate relevance. During training, GTRec and all LLM-based baselines sample hard negatives from the same SASRec-retrieved candidate pool to ensure fair comparison.
As shown in Table~\ref{tbl:topktable}, GTRec achieves the best overall performance on \textit{Steam} and \textit{LastFM} across both MLLM backbones in terms of HitRatio@5, HitRatio@10, NDCG@5, and NDCG@10. Compared with SASRec and LLM-based rerankers such as CoLLM and GLRec, GTRec leverages visual graph connectivity and CGC-compressed textual semantics to improve candidate relevance estimation and Top-$k$ ranking accuracy.

\begin{figure}[tbp]
    \centering
    \begin{minipage}{.49\linewidth}
        \centering
        \includegraphics[width=\linewidth]{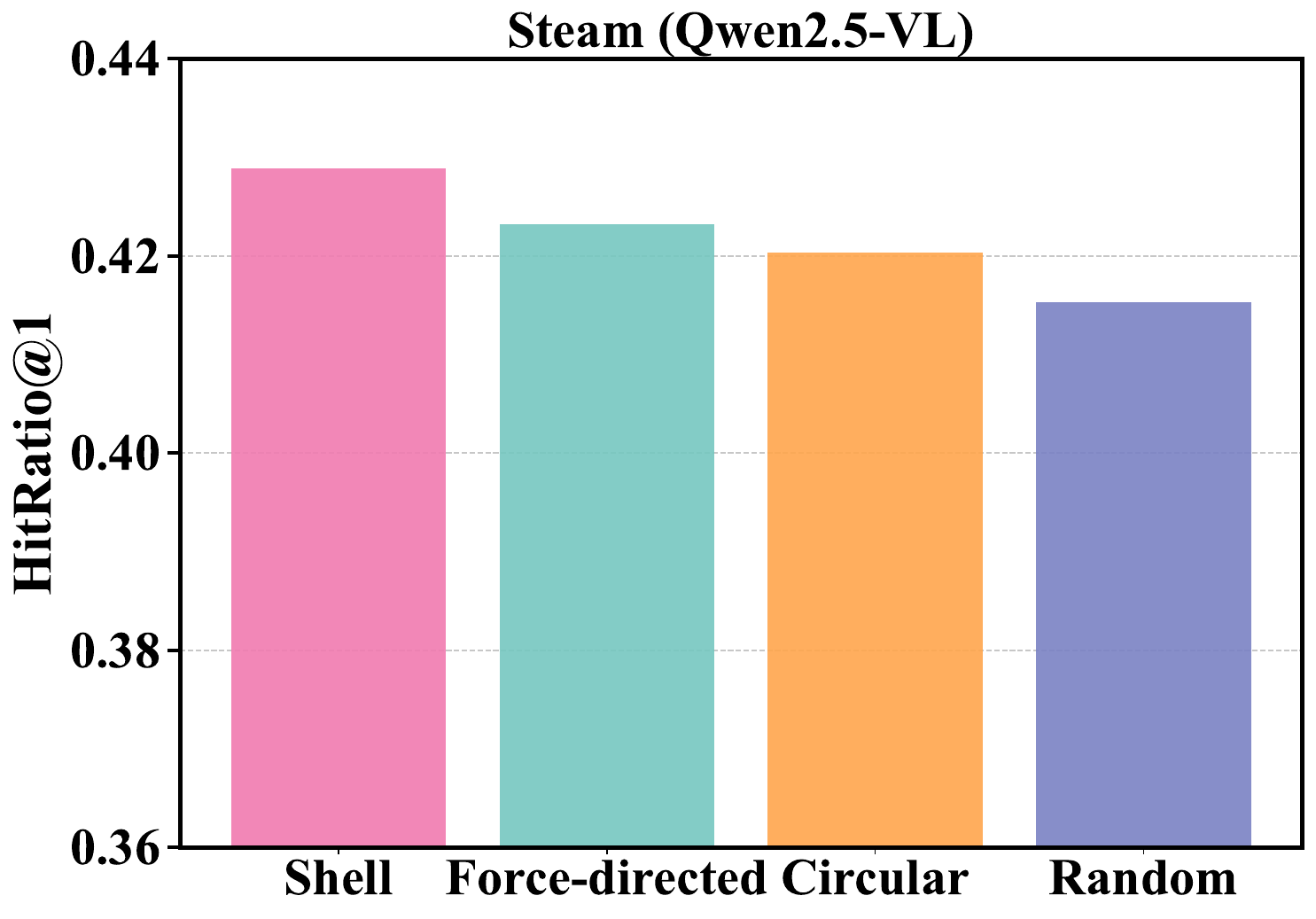}
    \end{minipage}\hfill
    \begin{minipage}{.49\linewidth}
        \centering
        \includegraphics[width=\linewidth]{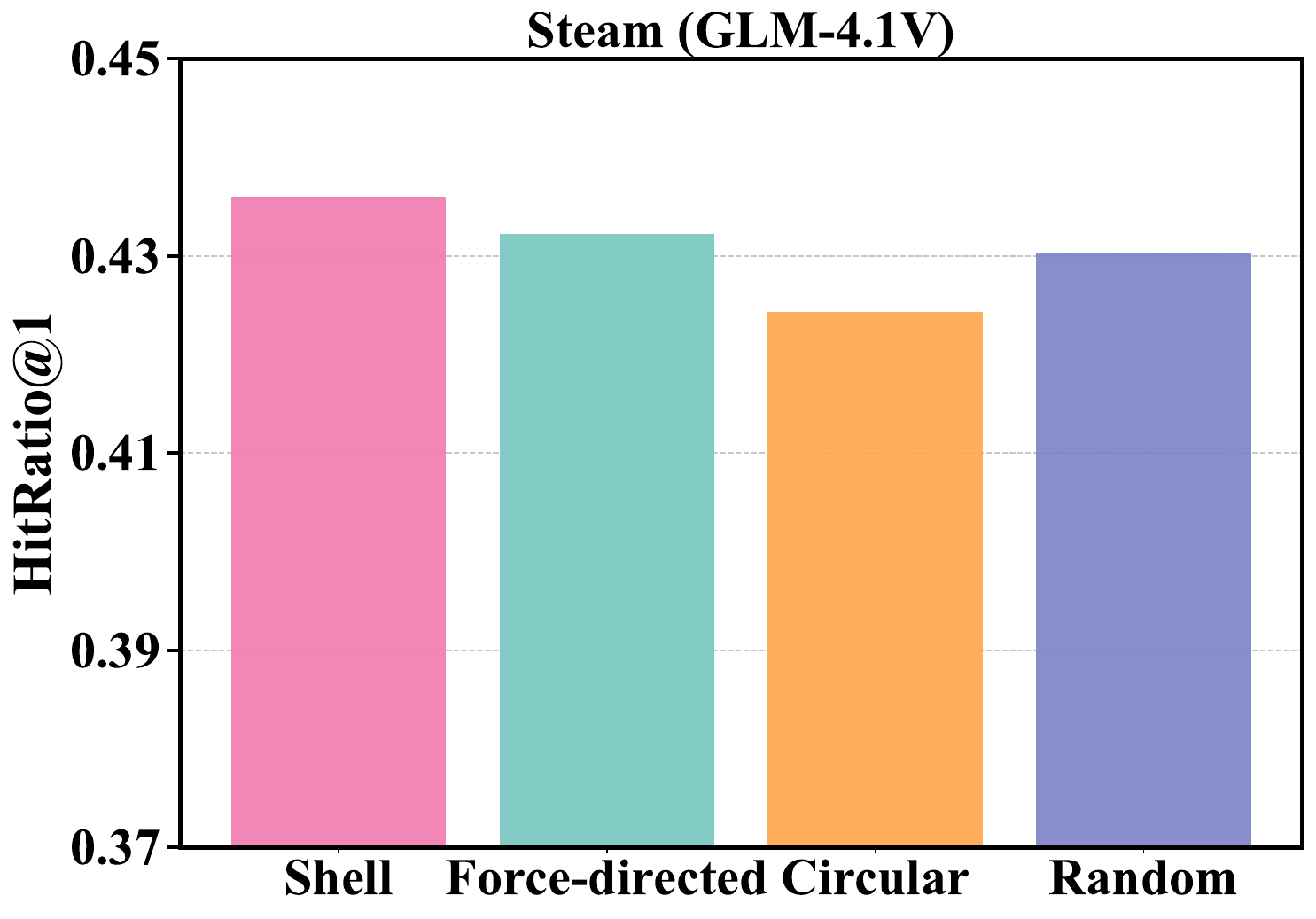}
    \end{minipage}
    \caption{HitRatio@1 of different graph layout strategies on the Steam dataset. }
    \label{fig:buju}
\end{figure}

\begin{table}[tbp]
\centering
\caption{Topology sensitivity analysis of visual MITG on \textit{Steam} and \textit{LastFM} with \textit{HitRatio@1} under the Qwen2.5-VL-7B-Instruct backbone.}
\setlength{\tabcolsep}{4pt}
\begin{tabular}{lcc}
\toprule
\textbf{Variants} & \textit{\textbf{Steam}} & \textit{\textbf{LastFM}} \\
\midrule
MITG & 0.4016 & 0.3924 \\
No-Edge      & 0.3453 ($\downarrow 14.02\%$) & 0.3271 ($\downarrow 16.64\%$) \\
Random-Edge  & 0.2970 ($\downarrow 26.05\%$) & 0.2856 ($\downarrow 27.22\%$) \\
No-Label     & 0.0643 ($\downarrow 83.99\%$) & 0.0527 ($\downarrow 86.57\%$) \\
\bottomrule
\end{tabular}
\label{tab:topology-sensitivity-qwen}
\end{table}

\subsection{Position-wise Information Utilization}

\begin{table*}[t]
\centering
\caption{Position-wise information utilization under controlled masking on the Steam dataset with GLM-4.1V-Thinking and Qwen2.5-VL. Higher relative performance degradation ($\Delta$) indicates stronger utilization of information at the corresponding position.}
\label{tab:position_masking}
\renewcommand{\arraystretch}{1.1}
\setlength{\tabcolsep}{6pt}
\begin{tabular}{ll|c|cc|cc|cc}
\toprule[1.2pt]
\textbf{Model}
& \textbf{Method}
& \textbf{Full}
& \textbf{w/o Head}
& $\boldsymbol{\Delta_H}$ \textbf{(\%) $\uparrow$}
& \textbf{w/o Middle}
& $\boldsymbol{\Delta_M}$ \textbf{(\%) $\uparrow$}
& \textbf{w/o Tail}
& $\boldsymbol{\Delta_T}$ \textbf{(\%) $\uparrow$} \\
\midrule
\multirow{3}{*}{Qwen2.5-VL}
& Linearized MITG
& 0.3825
& 0.2838 & 25.80
& 0.3164 & 17.28
& 0.2806 & 26.64 \\
& Visualized MITG
& 0.4016
& 0.3283 & 18.25
& 0.3219 & 19.85
& 0.3307 & 17.65 \\
& GTRec
& 0.4289
& 0.3463 & 19.26
& 0.3480 & 18.86
& 0.3506 & 18.26 \\
\midrule
\multirow{3}{*}{GLM-4.1V}
& Linearized MITG
& 0.3883
& 0.3035 & 21.84
& 0.3312 & 14.71
& 0.2943 & 24.21 \\
& Visualized MITG
& 0.4107
& 0.3530 & 14.05
& 0.3503 & 14.71
& 0.3465 & 15.63 \\
& GTRec
& 0.4360
& 0.3666 & 15.92
& 0.3650 & 16.28
& 0.3693 & 15.30 \\
\bottomrule[1.2pt]
\end{tabular}
\end{table*}

To examine positional information utilization across MITG representations, we conduct a controlled masking experiment on Steam. We divide the input into head, middle, and tail regions, randomly masking an equal proportion of information in each. Regional contributions are measured by relative performance degradation $\Delta_p=(P_{\mathrm{Full}}-P_{\mathrm{w/o},p})/P_{\mathrm{Full}}\times100\%$, where $p\in\{H,M,T\}$ and larger values indicate greater information utilization.
As shown in Table~\ref{tab:position_masking}, Linearized MITG exhibits positional imbalance across both models, with middle-region masking consistently causing smaller performance drops than head- or tail-region masking, indicating the lost-in-the-middle phenomenon. In contrast, Visualized MITG yields more balanced degradation across regions, while GTRec maintains this balance with the highest overall performance. These findings suggest that MITG visualization alleviates positional bias from sequential linearization and promotes more uniform information utilization.

\begin{table}[t]
\centering
\caption{Structural relation recovery on the Steam dataset.}
\label{tab:structural_recovery}
\renewcommand{\arraystretch}{1.25}
\begin{tabular*}{\columnwidth}{@{\extracolsep{\fill}} l l c c c}
\hline
Model & Method & 1-hop & 2-hop & 3-hop \\
\hline
\multirow{2}{*}{Qwen2.5-VL}
& Linearized MITG & 0.9623 & 0.8547 & 0.6835 \\
& Visualized MITG & 0.9634 & 0.8776 & 0.7321 \\
\hline
\multirow{2}{*}{GLM-4.1V}
& Linearized MITG & 0.9717 & 0.8562 & 0.7048 \\
& Visualized MITG & 0.9703 & 0.8815 & 0.7563 \\
\hline
\end{tabular*}
\end{table}

\subsection{Structural Fidelity Analysis.}
To directly evaluate the structural fidelity of different MITG representations, we construct a multi-hop structural relation recovery task. Given an MITG representation, the MLLM is required to determine whether a queried structural relation between two nodes holds at a specified hop distance. We report the recovery accuracy for 1-hop, 2-hop, and 3-hop relations on the Steam dataset.
As shown in Table~\ref{tab:structural_recovery}, Linearized MITG and Visualized MITG achieve comparable accuracy on 1-hop relations, indicating that both representations allow the MLLM to effectively capture local connectivity. However, the performance gap becomes more pronounced as the relation depth increases. For Qwen2.5-VL, Visualized MITG improves the 2-hop and 3-hop recovery accuracy from 0.8547 and 0.6835 to 0.8776 and 0.7321, respectively. A similar trend is observed for GLM-4.1V, where the corresponding accuracy increases from 0.8562 and 0.7048 to 0.8815 and 0.7563, respectively. These results indicate that sequential linearization makes it increasingly difficult for the MLLM to capture higher-order structural relations as the relation depth increases. In contrast, Visualized MITG enables more effective recovery of multi-hop topological relations, demonstrating that the visualized representation helps the MLLM better capture higher-order structural information.

\subsection{Effect of Hyperparameters}

\subsubsection{Impact of Graph Layout}
We investigate the effect of different graph layout strategies on GTRec’s recommendation performance. As shown in Fig.~\ref{fig:buju}, the Shell layout achieves the best performance on both Qwen2.5-VL and GLM-4.1V, indicating that concentric node organization better preserves the hierarchical and centrality information in the visual MITG. The Force-directed layout captures relational clustering well but reflects hierarchical structure less effectively. In contrast, Circular and Random layouts achieve lower HitRatio@1 due to the lack of clear structural information. Overall, GTRec remains robust across different layouts, while a well-designed layout can further improve recommendation performance.

\subsubsection{Topology Sensitivity Analysis}
To verify whether the MLLM exploits the relational topology in the visual MITG, we construct several controlled variants while keeping the user history, candidates, prompt template, image resolution, and node number unchanged. Specifically, \textit{No-Edge} removes all edges; \textit{Random-Edge} randomly rewires edges while preserving their number and layout; and \textit{No-Label} removes node titles.
As shown in Table~\ref{tab:topology-sensitivity-qwen}, \textit{No-Edge} and \textit{Random-Edge} lead to clear performance drops, indicating that the connectivity structure in visual MITG provides useful structural signals for recommendation. Removing node labels causes a much larger degradation, suggesting that node semantics are necessary for grounding the graph to user histories and candidate items. Overall, these results demonstrate that visual MITG benefits from the joint modeling of node semantics and graph connectivity.

\subsubsection{Image Resolution Sensitivity}
\begin{figure}[tbp]
    \centering
    \begin{minipage}{.49\linewidth}
        \centering
        \includegraphics[width=\linewidth]{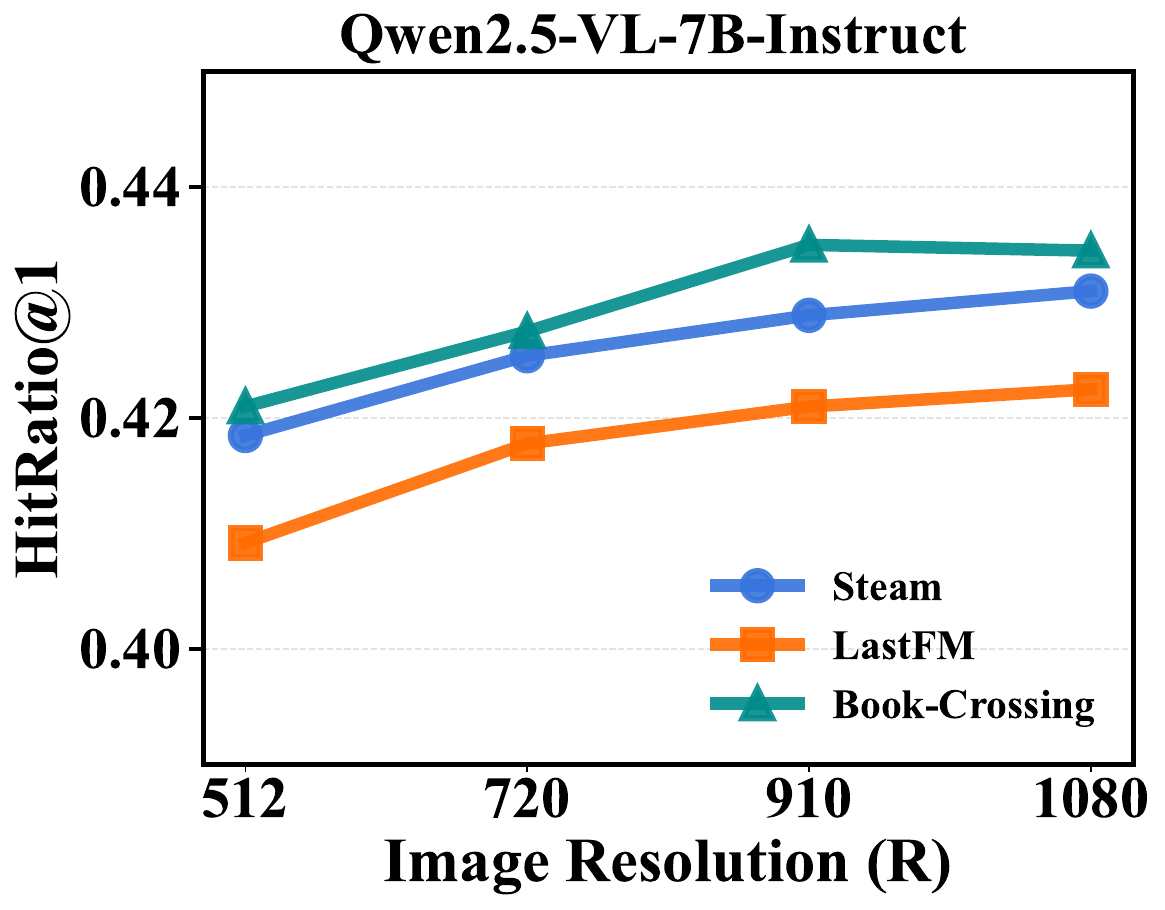}
    \end{minipage}\hfill
    \begin{minipage}{.49\linewidth}
        \centering
        \includegraphics[width=\linewidth]{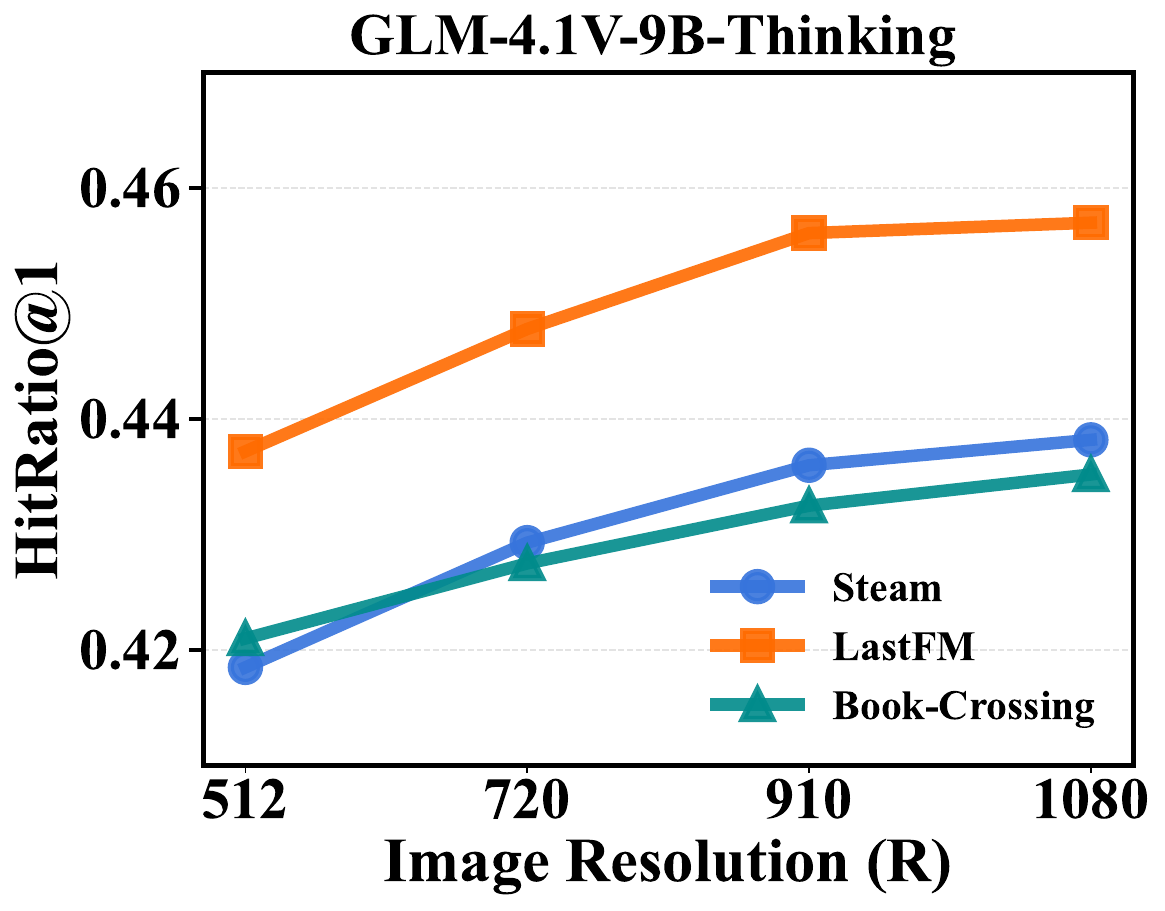}
    \end{minipage}
    \caption{Performance comparison of GTRec under different image resolutions on three datasets. }
    \label{fig:resolution}
\end{figure}

We further investigate the impact of image resolution in the visual transformation process.
As shown in Fig.~\ref{fig:resolution}, increasing the resolution from 512 to 910 consistently improves the \textit{HitRatio@1}, indicating that higher resolutions enable the model to capture more fine-grained topological structures within the visualized MITG.
However, further enlarging the resolution to 1080 yields only marginal accuracy gains while significantly increasing inference latency and GPU memory consumption.
Therefore, setting $R=910$ achieves an effective trade-off between structural fidelity and computational efficiency.

\subsubsection{Effect of Markov Step $k$}
\begin{figure}[tbp]
    \centering
    \begin{minipage}{.49\linewidth}
        \centering
        \includegraphics[width=\linewidth]{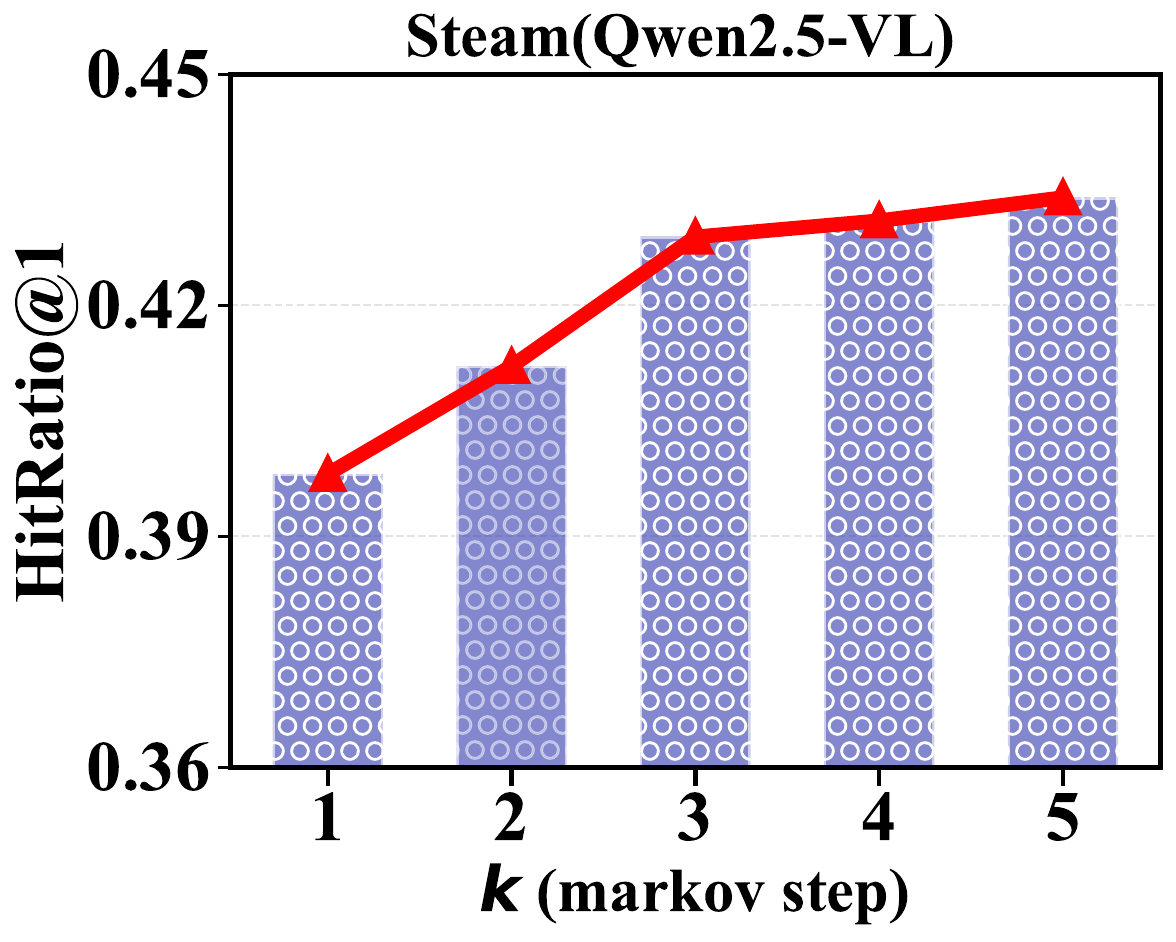}
    \end{minipage}\hfill
    \begin{minipage}{.49\linewidth}
        \centering
        \includegraphics[width=\linewidth]{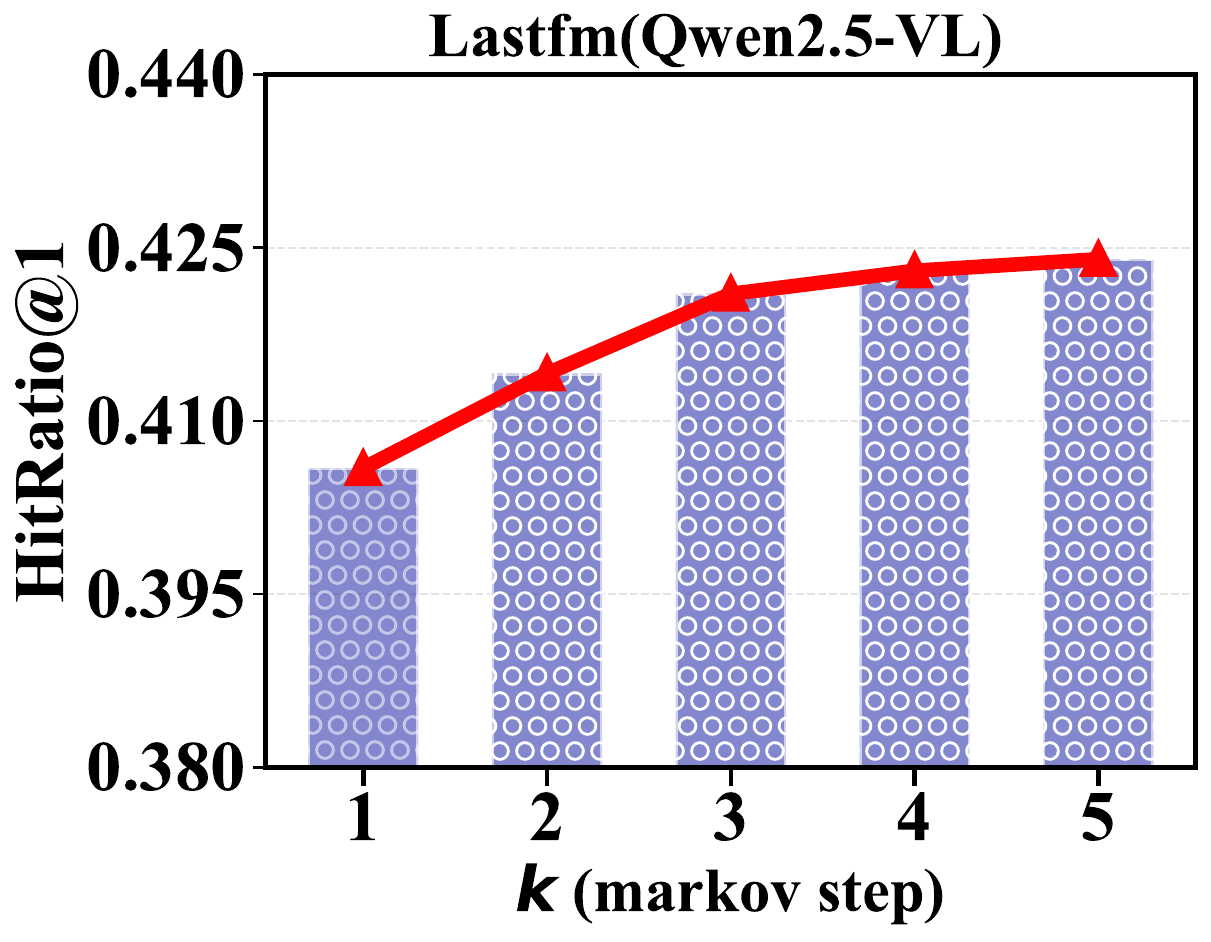}
    \end{minipage}
    \caption{Performance comparison of GTRec under different Markov step $k$.}
    \label{fig:markov}
\end{figure}

In the hyper-parameter sensitivity study, we analyze the impact of the Markov step $k$ in modeling short-term interests on model performance.
As shown in Fig.~\ref{fig:markov}, results show that as $k$ increases from 1 to 5, metrics \textit{HitRatio@1} improve initially and then plateau across both datasets. The best or near-best performance is consistently achieved when $k{=}3$. Smaller $k$ values fail to capture mid-range sequential dependencies, while larger $k$ values increase computational cost, leading to diminishing returns. Overall, $k{=}3$ strikes the best balance between accuracy and efficiency, and is thus used as the default setting.

\subsection{Training and Inference Efficiency}

We further evaluate the computational efficiency of Qwen3-VL and GLM-4.1V models on LastFM using a single NVIDIA PRO 6000 GPU, comparing textual descriptions with visualized graphs at a resolution of $910\times910$. As shown in Table~\ref{tab:qwen3vl_efficiency},
visualized graphs introduce additional training and inference overhead across all backbones, with varying increases in GPU memory consumption.
Notably, the relative inference overhead is lower for
Qwen3-VL-32B than for Qwen3-VL-8B, while the additional
GPU memory consumption is approximately 4.1\% and 1.2\%. 
Overall, although visualized graph representations introduce additional computational overhead, the increase in GPU memory
consumption remains relatively small for the Qwen3-VL backbones. Combined with the observed improvements in recommendation performance, these results demonstrate that visualized graph representations can effectively enhance recommendation accuracy without an excessive increase in computational cost, further highlighting the practical potential of our method in real-world recommendation tasks.

\begin{table}[tbp]
\centering
\caption{Computational efficiency of Qwen3-VL models on LastFM using a single NVIDIA PRO6000 GPU.}
\label{tab:qwen3vl_efficiency}
\small
\setlength{\tabcolsep}{4pt}
\resizebox{\columnwidth}{!}{%
\begin{tabular}{llccc}
\toprule
Model & Input & Train (s) & VRAM (GiB) & Test (s) \\
\midrule
\multirow{2}{*}{Qwen3-VL-8B}
& Textual Description & 0.320 & 17.593 & 0.146 \\
& Visualized Graph & 0.350 & 18.318 & 0.206 \\
\midrule
\multirow{2}{*}{Qwen3-VL-32B}
& Textual Description & 0.441 & 64.340 & 0.368 \\
& Visualized Graph & 0.642 & 65.140 & 0.497 \\
\midrule
\multirow{2}{*}{GLM-4.1V}
& Textual Description & 0.227 & 22.393 & 0.240 \\
& Visualized Graph & 0.268 & 33.949 & 0.349 \\
\bottomrule
\end{tabular}%
}
\end{table}

\section{CONCLUSION AND FUTURE WORK}
In this paper, we propose GTRec, a framework designed to mitigate the ``Lost-in-the-Middle'' problem and semantic flattening that limit existing ``graph-to-text'' LLM-based recommendation methods. We first construct the MITG to integrate collaborative and semantic knowledge. Then, we preserve global structural information through graph transformation. Furthermore, we design CGC to reduce input length while retaining essential semantics. By combining the visualized MITG and compressed textual semantics into a unified multimodal prompt, GTRec enables MLLMs to simultaneously capture structural dependencies and rich semantic information. Extensive experiments on multiple benchmark datasets demonstrate the effectiveness of our approach. In our paper, although our multimodal inference improves performance, it may introduce additional latency and computational overhead. In future work, we will explore lightweight modeling and inference acceleration, together with compression techniques such as distillation and quantization, to improve the efficiency of GTRec with minimal loss in accuracy.

\section*{Acknowledgment}
This research was supported by the National Natural Science Foundation of China (62402093, 62672090), the Sichuan Science and Technology Program (2025ZNSFSC0479), and the Fundamental Research Funds for the Central Universities (JBK202511020). This work was also supported by the Center for HPC, University of Electronic Science and Technology of China.

\ifCLASSOPTIONcaptionsoff
  \newpage
\fi

\bibliographystyle{IEEEtran}
\bibliography{sample-base}
\end{document}